\documentclass[article]{aa}  
\usepackage{graphicx}
\usepackage{txfonts}
\usepackage{xcolor}
\DeclareUnicodeCharacter{2212}{-}

\usepackage[]{hyperref}
\begin{document}

   \title{At the end of cosmic noon: Short gas depletion times in unobscured quasars at $z \sim$ 1}

   \author{
        M. Frias Castillo \inst{1}
        \and M. Rybak \inst{2, 1}
        \and J. Hodge\inst{1}
        \and P. van der Werf \inst{1}
        \and L. J. Abbo\inst{1}
        \and F. J. Ballieux\inst{1}
        \and S. Ward \inst{3,4,5}
        \and C. Harrison \inst{6}
        \and G. Calistro Rivera \inst{3}
        \and J. P. McKean \inst{7,8,9,10}
        \and H. R. Stacey \inst{3}
          \fnmsep
          }

   \institute{Leiden Observatory, Leiden University, Niels Bohrweg 2, 2333 CA Leiden, the Netherlands\\
             \email{friascastillom@strw.leidenuniv.nl}
   \and THz Sensing Group, Department of Microelectronics, Faculty of Electrical Engineering, Mathematics and Computer Science, Mekelweg 4, 2628 CD Delft, the Netherlands
   \and European Southern Observatory, Karl-Schwarzschild-Stra\ss e 2, D-85748 Garching bei M\"unchen, Germany 
   \and Excellence Cluster ORIGINS, Boltzmannstra\ss e 2, D-85748 Garching bei M\"unchen, Germany 
   \and Ludwig-Maximilian-Universit\"at, Professor-Huber-Platz 2, D-80539 M\"unchen, Germany 
   \and School of Mathematics, Statistics and Physics, Newcastle University, NE1 7RU, UK
   \and ASTRON, Netherlands Institute for Radio Astronomy, Oude Hoogeveensedijk 4, NL-7991 PD Dwingeloo, the Netherlands
   \and Kapteyn Astronomical Institute, University of Groningen, P.O. Box 800, 9700AV Groningen, the Netherlands 
   \and South African Radio Astronomy Observatory (SARAO), P.O. Box 443, Krugersdorp 1740, South Africa
   \and Department of Physics, University of Pretoria, Lynnwood Road, Hatfield, Pretoria 0083, South Africa}
   

  \abstract
   {Unobscured quasars (QSOs) are predicted to be the final stage in the evolutionary sequence from gas--rich mergers to gas--depleted, quenched galaxies.
   Studies of this population, however, find a high incidence of far--infrared--luminous sources---suggesting significant dust-obscured star formation---but direct observations of the cold molecular gas fuelling this star formation are still necessary. 
   We present a NOEMA study of CO(2--1) emission, tracing the cold molecular gas, in ten lensed $z=$1--1.5 unobscured QSOs. We detected CO(2--1) in seven of our targets, four of which also show continuum emission ($\lambda_\mathrm{rest}$ = 1.3mm). After subtracting the foreground galaxy contribution to the photometry, spectral energy distribution fitting yielded stellar masses of 10$^{9-11}$ M$_\odot$, with star formation rates of 25--160 M$_\odot$ yr$^{-1}$ for the host galaxies. These QSOs have lower $L'_\mathrm{CO}$ than star--forming galaxies with the same $L_\mathrm{IR}$, and show depletion times spanning a large range (50--900 Myr), but with a median of just 90($\alpha_\mathrm{CO}$/4) Myr. We find molecular gas masses in the range $\le$2--40 $\times$ 10$^9$($\alpha_\mathrm{CO}$/4) M$_\odot$, which suggest gas fractions above $\sim$50\% for most of the targets. Despite the presence of an unobscured QSO, the host galaxies are able to retain significant amounts of cold gas. However, with a median depletion time of $\sim$90 Myr, the intense burst of star formation taking place in these targets will quickly deplete their molecular gas reservoirs in the absence of gas replenishment, resulting in a quiescent host galaxy. The non--detected QSOs are three of the four radio--loud QSOs in the sample, and their properties indicate that they are likely already transitioning into quiescence. Recent cosmological simulations tend to overestimate the depletion times expected for these $z\sim1$ QSO--host galaxies, which is likely linked to their difficulty producing starbursts across the general high-redshift galaxy population.  }

    \titlerunning{CO(2--1) in Lensed QSOs at $z\sim$1}
    \authorrunning{Frias Castillo et al.}
   \maketitle
   
%

\section{Introduction}

The nature of massive galaxies changes dramatically between the epoch of the peak star-forming activity of the Universe ($z =$ 1--3, `cosmic noon'), and the present day \citep[e.g.][]{casey2014}. At cosmic noon, the most massive galaxies are typically sub-millimetre galaxies (SMGs) -- gas-rich systems with star formation rates (SFR) of 10$^2$ – 10$^3$ M$_{\odot}$ yr$^{-1}$. These are likely progenitors of the $z =$ 0 early type, quiescent galaxies. According to one of the most popular models of massive galaxy evolution, these changes are driven by interactions and mergers of gas--rich SMGs \citep[e.g.][]{sanders1988,alexander2005,hopkins2008,page2012,somerville2015}. Following the merger, gas is funnelled to the centre of the galaxy through the loss of angular momentum, triggering powerful active galactic nuclei (AGN), accompanied by a starburst \citep{hopkins2006}. The dusty interstellar medium (ISM) results in an obscured quasar (QSO). Eventually, feedback from the QSO and/or star formation drives the cold gas and dust out of the galaxy through winds and outflows, quenching star formation in the host galaxy and becoming an unobscured QSO. This scenario naturally explains the coevolution of the black hole (BH) and its host galaxy \citep{kormendy_ho2013}, which is implied by the observed relation between the mass of the BH and the mass and velocity dispersion of their host spheroids \citep{ferrarese2000,gebhardt2000}.

Quasars that are luminous in the far-infrared (FIR) to millimetre regime are thought to be a transition phase  between the FIR-bright SMG and FIR-faint, unobscured QSO phases, in which the QSO host galaxies have depleted gas reservoirs but are still able to sustain dust-obscured star formation. The low spatial density of FIR—luminous QSOs, compared to that of SMGs or UV--bright QSOs, previously led some studies to argue that this is a short—lived transition, with timescales as short as 1 Myr \citep{simpson2012}.  Far-infrared studies of QSOs consistently find, however, that they tend to reside in host galaxies with ongoing star formation \citep[e.g.][]{gurkan2015,harris2016,netzer2016}, with SFRs similar to those of normal, star-forming \citep[e.g.][]{rosario2013,stanley2017} or even gas--rich, starbursting galaxies \citep[e.g.][]{pitchford2016}. A \textit{Herschel}/SPIRE snapshot imaging survey of all high--redshift lensed QSOs known at the time by \citet{Stacey2018} revealed that $\sim$70\% of QSOs still have significant obscured star formation. This high incidence of an active AGN and high levels of star formation suggests that FIR—luminous QSOs might be able to maintain star formation for longer periods of time, of the order of $\sim$100 Myr, despite the presence of feedback from ongoing AGN activity. 

Studies of the cold ISM in these systems are crucial to understand what drives the high SFRs. Traced by the rotational emission lines of the carbon monoxide molecule (CO), the cold molecular gas is a fundamental ingredient of the ISM of galaxies, as it directly fuels both star formation \citep{kennicutt1998,bigiel2008} and accretion onto supermassive BHs \citep[see ][for a review]{storchi2019}. Results from studies at $z\ge$2 present discrepant findings. Some studies find that QSO host galaxies have lower gas fractions and gas depletion times than their non-QSO counterparts \citep{kakkad2017,perna2018,bischetti2021,circosta2021} and interpret this as highly efficient gas consumption due to the AGN feedback affecting the gas. These studies are, however, limited to the most massive and brightest systems due to sensitivity limitations. Other studies of AGN host-galaxies find that they have gas fractions that are indistinguishable from normal star-forming galaxies (SFGs) on the main sequence \citep{rodighiero2019, valentino2021}. In the local Universe, meanwhile, studies find no evidence of the impact of AGN feedback, with the gas reservoirs of AGN--host galaxies showing similar properties to those of SFGs \citep{husemann2017,rosario2018,jarvis2020,shangguan2020b,yesuf2020,koss2021,zhuang2021}.

The redshift $1 < z < 1.5$ corresponds to the end of the peak epoch of both star formation and accretion activity of BHs \citep{shankar2009,madau_dickinson2014}, and is therefore a crucial laboratory to look for AGN feedback effects. There is little direct knowledge (via CO observations) of the molecular gas reservoirs in FIR—luminous QSOs at this redshift range, right at the end of cosmic noon, when the energy input into the host galaxy from the AGN might be maximised. Previous work at $z=1-1.5$ was undertaken as part of a larger CO survey of 18 gravitationally lensed QSOs in the redshift range $\sim$1.3--3.8 by \cite{barvainis2002-co}, who obtained only upper limits at $z\sim1.5$. To date, there have only been two CO detections of unobscured, FIR—bright QSOs, which should be about to enter the quenched phase of their evolutionary path. These are the lensed  Q 0957+561 \citep{krips2005} and HS 0810+2554 \citep{chartas2020,stacey2021}, which have molecular gas masses of $\simeq$10$^{10}$ and $\simeq$ 3-5$\times$10$^9$ M$_\odot$, respectively, with estimated depletion times of $\simeq$100 Myr. At a slightly lower redshift, however, the strongly lensed QSO RXJ1131-1231 ($z\sim$0.65) has a massive ($\sim$10$^{11}$ M$_{\odot}$) molecular gas disc $\sim$15 kpc in diameter \citep{Paraficz2018}, with a depletion time of 1 Gyr, more typical of normal SFGs \citep{tacconi2018}. The large difference in molecular gas mass and depletion times suggests a large scatter in cold gas content among unobscured QSOs at $z\sim$1—1.5.

\begin{table*}[t]
\caption{Target sample, ordered by increasing source redshift. Columns give the following information: target name, celestial coordinates, redshift, lens redshift, IR luminosities, IR-based SFR,  intrinsic luminosity at 3000 \AA and MgII-based BH mass. The last two are adopted from \citet{Stacey2018} and are not corrected for the lensing magnification. \label{tab:sample} }
\hspace{-0.5cm}
\begin{tabular}{@{}lccccccccc @{}}
 \hline \hline
Target & RA & Dec & $z_S^\mathrm{opt}$ & $z_L$ & log $\mu_\mathrm{IR} L_\mathrm{IR}$ & log $\mu_\mathrm{IR}$SFR & $\mu_\mathrm{IR}$ & log($\mu L_{3000}$)$^*$ & log($\mu M^{MgII}_\mathrm{BH}$)$^*$\\
& \multicolumn{2}{c}{J2000} & & & [$L_\odot$] & [$M_\odot$ yr$^{-1}$] & [erg s$^{-1}$] [$M_\odot$]\\
\hline
J1524+4409 & 15:24:45.63 & 44:09:49.6 & 1.211 & 0.320 & 12.2$^{+0.3}_{-0.2}$ & 2.7$^{+0.3}_{-0.2}$ & 7.4$^a$ & 45.585$\pm$0.005 & 9.54$\pm$0.20\\
B1608+656$^{**}$ & 16:09:13.96 & 65:32:29.0 & 1.394 & 0.630&  11.9$^{+0.2}_{-0.1}$ & 2.4$^{+0.2}_{-0.1}$ & 10.8$^b$ & -- & -- \\
J1330+1810 & 13:30:18.65 & 18:10:32.1 & 1.394 & 0.373&  12.8$^{+0.1}_{-0.1}$ & 3.3$^{+0.1}_{-0.1}$& 24$\pm1^c$ & 45.969 $\pm$ 0.006 & 9.19$\pm$0.02\\
J1455+1447 & 14:55:01.91 & 14:47:34.8 & 1.426 & 0.42&  12.6$^{+0.3}_{-0.2}$ & 3.1$^{+0.3}_{-0.2}$ & 12.7$\pm2^d$ & 45.944 $\pm$0.002 & 8.46$\pm$0.03\\
J1633+3134$^{**}$ & 16:33:49.00 & 31:34:12.0 & 1.523 & 0.684 & 12.9$^{+0.2}_{-0.2}$ &3.5$^{+0.2}_{-0.2}$ & 10$\pm$5 &  46.439$\pm$0.002 & 9.27$\pm$0.03\\
J0924+0219 & 09:24:55.83 & 02:19:23.6 & 1.525 & 0.393&  12.5$^{+0.2}_{-0.1}$ & 3.0$^{+0.2}_{-0.1}$& 17$\pm1^e$ & 45.970 $\pm$ 0.002 & 8.76$\pm$0.11\\
J1650+4251 & 16:50:43.33 & 42:51:49.3 & 1.543 & 0.577 &12.5$^{+0.1}_{-0.1}$ &3.0$^{+0.1}_{-0.1}$& 10$\pm$5 & 46.391$\pm$0.001 & 9.60$\pm$0.02\\
B1152+200$^{**}$ & 11:55:18.29 & 19:39:42.0 & 1.019 & 0.438 &  12.2$^{+0.8}_{-0.3}$ & 2.7$^{+0.8}_{-0.3}$ & 10$\pm$5 & 46.062$\pm$0.001 & 9.73$\pm$0.02\\
B1600+434$^{**}$ & 16:01:40.48  & 43:16:47.4 & 1.59 &  0.414&  12.4$^{+0.2}_{-0.2}$ & 2.9$^{+0.2}_{-0.2}$& 10$\pm$5 & -- & -- \\
J0806+2006 &  08:06:23.70 & 20:06:31.8 & 1.542 & 0.573 &  12.4$^{+0.4}_{-0.2}$ & 2.9$^{+0.4}_{-0.2}$& 10$\pm$5 & 45.981$\pm$0.002 & 8.70$\pm$0.10 \\
\hline
 \hline
 \multicolumn{9}{l}{$^*$ Taken from \cite{rakshit2020}. B1152+656 and B1600+434 do not have available SDSS spectra.}\\
 \multicolumn{9}{l}{$^{**}$Sources are classified as radio--loud, `jetted' QSOs  \citep{Stacey2018}}\\
 \multicolumn{9}{l}{References: $^a$\cite{oguri2008}, $^b$\cite{barvainis2002}, $^c$\cite{stacey2022},$^d$\cite{kayo2010}, $^e$\cite{stacey2021}}\\
 \end{tabular}
\end{table*}

Motivated by these findings, we have targeted the CO(2--1) ($\nu_\mathrm{rest}$=230.5380 GHz) emission line in a sample of strongly lensed unobscured QSOs in the redshift range $z =$ 1--1.5 using the NOrthern Extended Millimitre Array (NOEMA). These are all the QSOs in this redshift range which are detected in targeted \textit{Herschel} SPIRE photometry from the survey by \cite{Stacey2018} and that are observable by NOEMA. By targeting gravitationally lensed objects, we probe fainter systems that would otherwise need prohibitively long integration times to be detected, although this comes at the expense of differential magnification effects. This survey aims to study the gas content in these intermediate--redshift QSOs, establishing their gas depletion times and gas fractions. With this data, we aim to fit these QSOs in the canonical SMG-QSO evolution scenario. Specifically, we aim to answer the question of whether their gas reservoirs are massive enough to maintain their obscured SFRs for long periods of time, or they are about to be quenched. 

The paper is structured as follows. In Section \ref{sec:observations} we describe the sample selection, NOEMA observations, and data reduction. In Section \ref{sec:results}, we present the CO(2--1) line and continuum detections, a search for the HCN/HNC/HCO$^+$ lines covered by our observations, and we describe our SED fitting approach. In Section \ref{sec:Analysis} we analyse the $L'_\mathrm{CO}-L_\mathrm{IR}$ relation for the sample (\ref{sec:Lir}), their total cold molecular gas content(\ref{sec:Lco}), gas fractions and depletion times (\ref{sec:tdep}), and compare our results with different studies from the literature, as well as state-of-the-art simulations (\ref{sec:sims}). Finally, we present our conclusions in Section \ref{sec:conclusions}. Throughout this paper, we adopt a flat universe model with a Hubble constant of H$_0$ = 67.8 km s$^{-1}$ Mpc$^{-1}$, $\Omega_M$ = 0.31, and $\Omega_\Lambda$ = 0.69 \citep{Planck2016}. 

\section{Observations and data reduction}
\label{sec:observations}

\subsection{Target sample}

Our targets are drawn from the sample of 104 gravitationally lensed QSOs observed with the \textit{Herschel Space Observatory} \citep{pilbratt2010-herschel} using the Spectral and Photometric Imaging Receiver (SPIRE) instrument \citep{griffin2010-spire}. We refer the reader to \cite{Stacey2018} for further details on the selection of the parent sample. From this list of targets, we selected all QSOs detected in FIR with known redshifts within the range $z=$1--1.5 for which the CO(2--1) emission line is observable with NOEMA. Three of our sources, B1608+656, B1152+200, and B1600+434, have strong jet--dominated radio emission \citep{browne2003}.

The final sample is shown in Table \ref{tab:sample}. The magnification factors listed have been derived from high-resolution observations in the FIR to sub-millimetre regime. When no magnification has been derived for a source, we assume a magnification of $\mu_\mathrm{FIR}$=10$^{+10}_{-5}$ for the intrinsic properties discussed throughout the paper and propagate the errors accordingly. We note that the bulk of our analysis is based on brightness ratios and thus independent of the magnification factor. 

A different source-plane distribution of the dust (FIR), gas (CO), and stellar emission might however result in a  differential magnification bias - that is, each tracer is magnified by a different factor. These can be significant, especially in highly magnified systems - for example, in the strongly lensed AGN host B1938+666, the FIR and CO(1--0) emission are magnified by a factor of $\approx$16 and $\approx9$, respectively \citep{Spingola2020}. While our spatially unresolved CO(2--1) observations do not allow us to derive the corresponding magnifications via lens modelling, we do not expect significant differential magnification bias in our sample. First, out of the seven sources detected in CO(2--1), five are doubly imaged, which reduces the differential magnification. Second, even in quadruply lensed systems such as SDP.81, the difference between FIR continuum and CO magnifications is $\leq$20\% \citep{Rybak2020b}, comparable to other uncertainties in our analysis (such as relative flux calibration). Therefore, we do not expect differential magnification to affect the main conclusions of this paper. Future resolved CO(2--1) and (sub)mm-wave imaging will be necessary to derive proper magnification factors for each component. 

\subsection{NOEMA observations and data reduction}

The observations were conducted as a part of the NOEMA projects S19CC and W20CM (PI: M. Rybak) between 2019 June 8 and 2020 December 30 in Band 1 in D configuration, with nine to ten 15-m antennas. The details of the observations are summarised in Table~\ref{tab:obs}. The water vapour (pwv) estimates are based on the 22-GHz radiometer measurements. At our targeted frequency (100~GHz), the primary beam has a full width at half-maximum (FWHM) of $\sim$50~arcsec. We used the PolyFiX correlator with a standard spectral resolution of 2 MHz to cover a total bandwidth of 15 GHz. 

The observations were tuned to centre the CO(2--1) transition of each source in the upper sideband. Depending on the target and time of observation, one of the two strong radio stars MWC349 or LKHA101 were observed for absolute flux calibration. For bandpass calibration, we used either 3C84, 3C279, 3C345, 1055+018, 1633+382, 1749+428 or 2013+370, depending on the target and time of observation. We integrated between 2.2 and 10.8 h on each source, resulting in a noise level in the range of 0.2--0.5 mJy beam$^{-1}$ at a spectral resolution of 20~MHz using natural weighting. The final integrated beam sizes and sensitivities are reported in Table \ref{tab:noema_beams}.

\begin{table}
\caption{NOEMA imaging: final synthesised beam sizes and sensitivity for the integrated continuum (given at 1.3 mm rest-frame) and 20~MHz bandwidth at the position of the CO(2--1) line ($\sim$60 km s$^{-1}$), for natural weighting. \label{tab:noema_beams} }
\begin{center}
 \begin{tabular}{@{}lcccc @{}}
 \hline \hline
Target & FWHM & PA & $\sigma_\mathrm{1.3mm}$ & $\sigma_\mathrm{20 MHz}$\\
& [arcsec] & [deg] & $\mu$Jy beam$^{-1}$ & mJy beam$^{-1}$\\
\hline
J1524+4409 &  4.2$\times$3.9 & -180 & 25 & 0.3  \\
B1608+656 & 6.0$\times$3.7 & 81 & 12 & 0.3 \\
J1330+1810 & 4.1$\times$2.3 & -169 & 25 & 0.2 \\
J1455+1447 & 4.3$\times$2.4 & -169 & 25 & 0.2 \\
J1633+3134 & 5.1$\times$3.9 & 20 & 14 & 0.2 \\
J0924+0219 & 3.4$\times$2.2 & 31 & 200 & 0.4 \\
J1650+4251 & 4.3$\times$3.6 & -6 & 16 & 0.3 \\
B1152+200 & 2.3$\times$2.2 & -15 & 17 & 0.5 \\
B1600+434 & 2.2$\times$2.0 & 57 & 70 & 0.2 \\
J0806+2006 & 3.3$\times$1.9 & 32 & 14 & 0.2 \\
\hline
 
 \hline
 \end{tabular}
\end{center}
\end{table}

Data calibration, cleaning, and imaging was carried out using the \textsc{Gildas} \textsc{Clic} software package\footnote{\texttt{http://www.iram.fr/IRAMFR/GILDAS}}. We selected channels on both sides of the CO(2--1) emission line as the fitting windows of linear baselines and subtracted a zeroth-order baseline from the cubes in the image plane to remove any continuum emission. We then imaged the residual spectral-line visibilities at spectral resolutions of 2--50 MHz. Finally, we also created maps of the 3 mm continuum emission after masking channels where the CO(2--1) line was detected. With the angular resolution achieved in this configuration (Table \ref{tab:noema_beams}), the targets are marginally resolved.

For B1608+656, B1152+200, and B1600+434, we used the very strong continuum signal (S/N$>$100) to self-calibrate the data, solving for the phase only. No CO(2--1) emission was detected before or after the self-calibration.

\begin{figure*}
    \centering
    \vspace{-0.4cm}
    \includegraphics[scale=0.6]{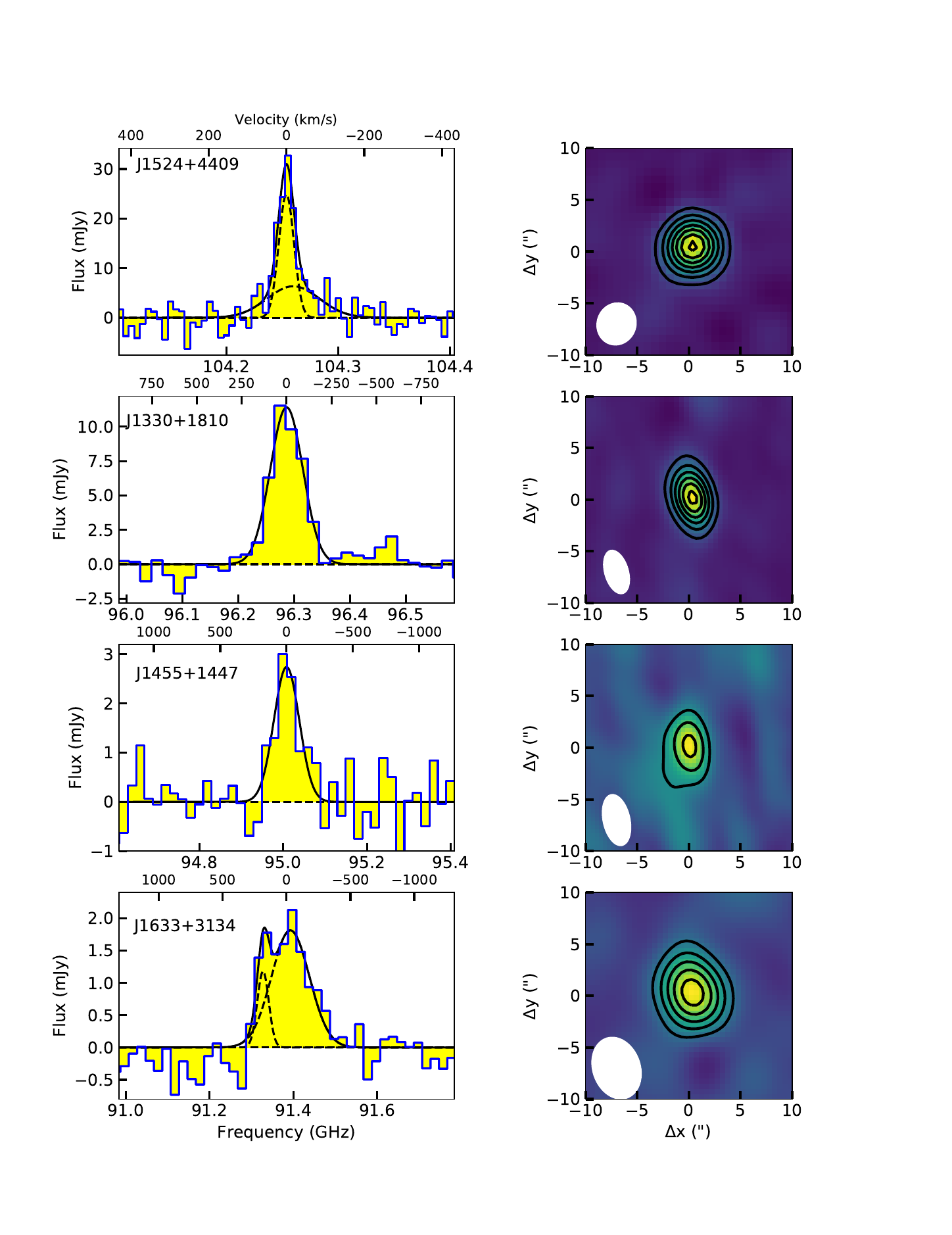}
    \caption{CO(2--1) spectra and 0th moment maps for the detected QSOs. The black solid line indicates the best fit Gaussian model, and the reference velocity was determined based on the redshift derived from the Gaussian fit (LSRK frame). Contours in the 0th moment maps start at 4$\sigma$ and increase in steps of 4$\sigma$ for J1524+2209, J1330+1810, J0924+0219, and J1650+4251; contours for J1455+1447, J1633+3134, and J0806+2006 start at 2$\sigma$ and increase in steps of 2$\sigma$. We have used the AIC criterion to determine whether a single- or double--Gaussian fit is more appropriate for the profiles.}
    \label{detections}
\end{figure*}

\begin{figure*}
    \centering
    \vspace{-0.4cm}
    \includegraphics[scale=0.6]{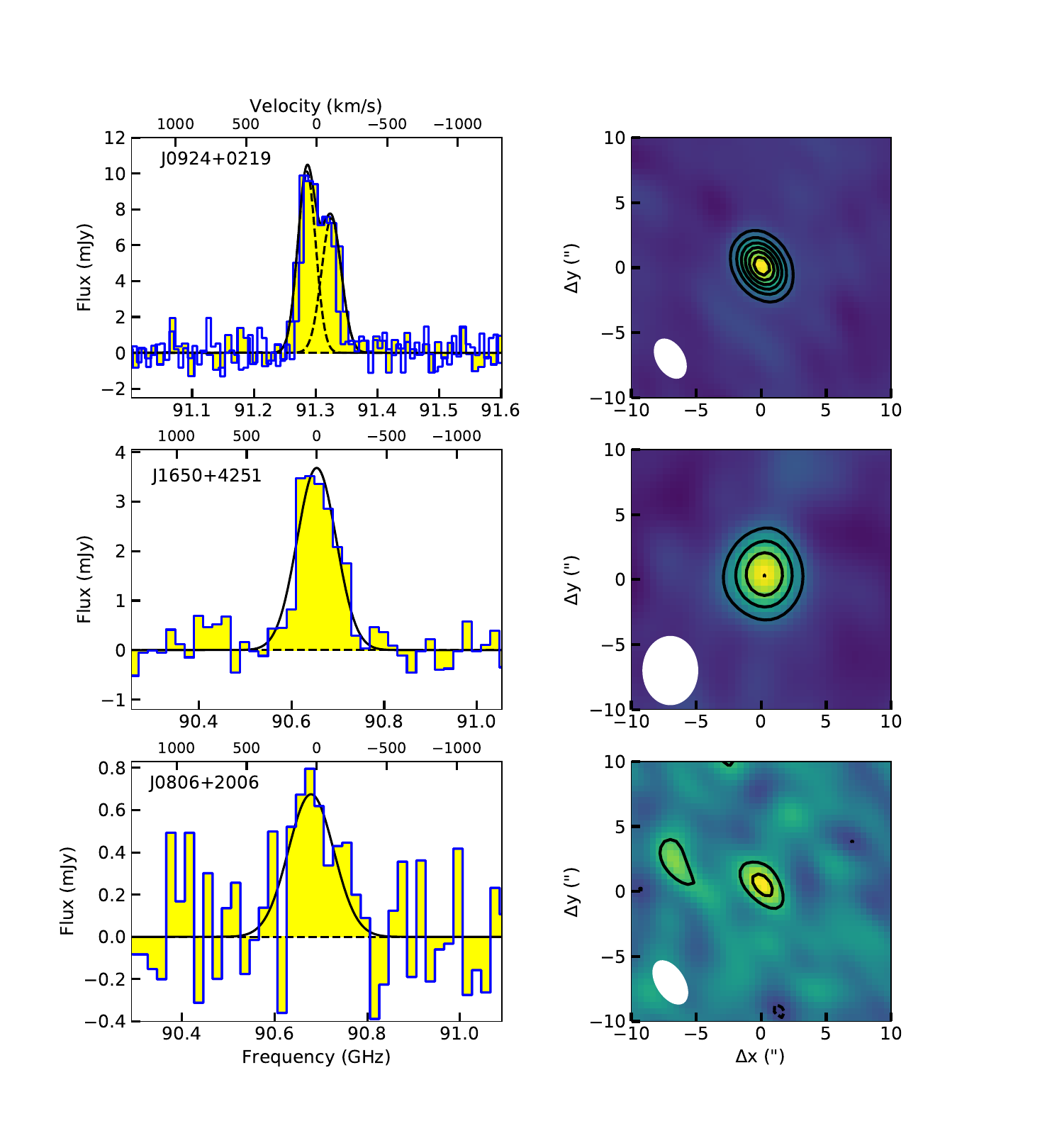}
    \caption{Fig. \ref{detections} - continued.}
    \label{detections_cont}
\end{figure*}

\section{Results} \label{sec:results}

\subsection{CO(2--1) line} \label{sec:detections}

To determine the flux and FWHM of the CO(2--1) line, we proceeded as follows. For each target, we considered the continuum-subtracted cube and extracted a line spectrum using a beam-sized aperture placed on the central pixel. The extracted spectrum was then fit with a one-dimensional Gaussian to determine the line centroid, $z_\mathrm{CO}$, and FWHM. We used the task IMMOMENTS in CASA \citep{casa} to create moment 0 maps by collapsing the spectral channels around the line centroid over a velocity range twice the CO(2--1) line FWHM for each target. The rms of the collapsed maps was estimated over an area approximately half that of the primary beam. We detected the CO(2--1) line in seven out of ten sources (Figures \ref{detections} and \ref{detections_cont}). Since not all our sources showed Gaussian profiles and their emission was extended beyond a beam, we extracted total line fluxes from the moment 0 maps instead of the Gaussian line fits to ensure we recovered all the flux. We did not find evidence of broad and/or asymmetric wings in the CO spectra that could indicate cold gas outflows.

\begin{table*}
\caption{Line and continuum detections: CO(2--1) line flux, line FWHM (measured from the spectra), CO(2-1) line luminosity, CO(1-0) line luminosity given $r_{21}=0.86$, and the rest-frame 1.3-mm continuum. Luminosities and fluxes were corrected for magnification as given in Table \ref{tab:sample}. Upper limits are quoted at the 3$\sigma$ level.  \label{tab:results} }
\begin{center}
 \begin{tabular}{@{}lcccccc @{}}
 \hline \hline
Target & z$_\mathrm{CO(2-1)}$ & $S_\mathrm{CO(2-1)} \Delta v$ & FWHM$_\mathrm{CO}$ & $L'_\mathrm{CO(2-1)}$ ($\times$10$^{9}$) & $L'_\mathrm{CO(1-0)}$ ($\times$10$^{9}$) & $S_\mathrm{1.3~mm}$ \\
 & & [Jy km s$^{-1}$] & [km s$^{-1}$] &  [K km s$^{-1}$ pc$^{-2}$] &  [K km s$^{-1}$ pc$^{-2}$] & [mJy]\\
\hline
J1524+4409 & 1.2113$\pm$0.0001 & 0.28$\pm$0.01 & 64$\pm$4 & 5.77$\pm$0.22 & 6.71$\pm$0.26 &  0.011$\pm$0.006 \\
B1608+656$^*$ & -- & $<$0.01 & -- & $<$0.3 & $<$0.4  & 1.7$\pm$0.06 \\
J1330+1810 & 1.3943$\pm$0.0002 & 0.08$\pm$0.01 & 214$\pm$12 & 2.26$\pm$0.15 & 2.63$\pm$0.17 &  0.002 $\pm$0.001\\
J1455+1447 &  1.4265$\pm$0.0005 & 0.05$\pm$0.01 & 224$\pm$31 & 4.54$\pm$0.31 & 1.74$\pm$0.37 &  $<$0.04\\
J1633+3134$^*$ & 1.5229$\pm$0.0003 &0.06$\pm$0.01 & 408$\pm$37 & 1.93$\pm$0.90 & 2.24$\pm$1.16 & 0.006 $\pm$ 0.002\\
J0924+0219 & 1.5249$\pm$0.0003 &0.14$\pm$0.01 & 216$\pm$12 & 4.53$\pm$0.31 & 5.27$\pm$0.36 &  $<$0.04 \\
J1650+4251 & 1.5431$\pm$0.0003 &0.10$\pm$0.01 & 341$\pm$19 & 3.21$\pm$1.60 & 3.73$\pm$1.90 & 0.011$\pm$0.002 \\
B1152+200$^*$ & -- & $<$0.03 & - & $<$0.4 & $<$0.5 & 1.35 $\pm$ 0.01\\
B1600+434$^*$ & -- & $<$0.01 & - & $<$0.4 & $<$0.5 & 3.43 $\pm$ 0.01 \\
J0806+2006 & 1.5422$\pm$0.0006 &0.03$\pm$0.01 & 548$\pm$ 36 & 0.87$\pm$0.49 & 1.01$\pm$0.57 & $<$0.004 \\
\hline
\multicolumn{6 }{l}{$^*$Radio--loud, `jetted' QSOs \citep{Stacey2018}}\\
  \hline \end{tabular}
\end{center}
\end{table*}

To determine the optimal mask for extraction of the line flux, we performed a curve of growth analysis. We iteratively extracted the flux from a circular aperture increased by 1$"$ at a time, until the extracted flux converged. We found that the flux was extended out to a radius of approximately 6$''$, so we chose this as the aperture radius to measure the final line fluxes. The line fluxes were consistent with those obtained if we fit a 2D Gaussian to the detected emission. For the sources without a CO detection, we provide a 3$\sigma$ upper limit, calculated from the rms of the velocity-integrated maps collapsed over a line width of 245 km s$^{-1}$, the median FWHM measured for our detections. The results of our analysis are reported in Table \ref{tab:results}. We measured integrated flux densities of the CO(2--1) line in the range $<$0.01--0.28 Jy km s$^{-1}$ (corrected for magnification) and FWHM in the range 65--550 km s$^{-1}$.

\begin{figure*}[!htb]
    \centering
    \includegraphics[scale=0.7]{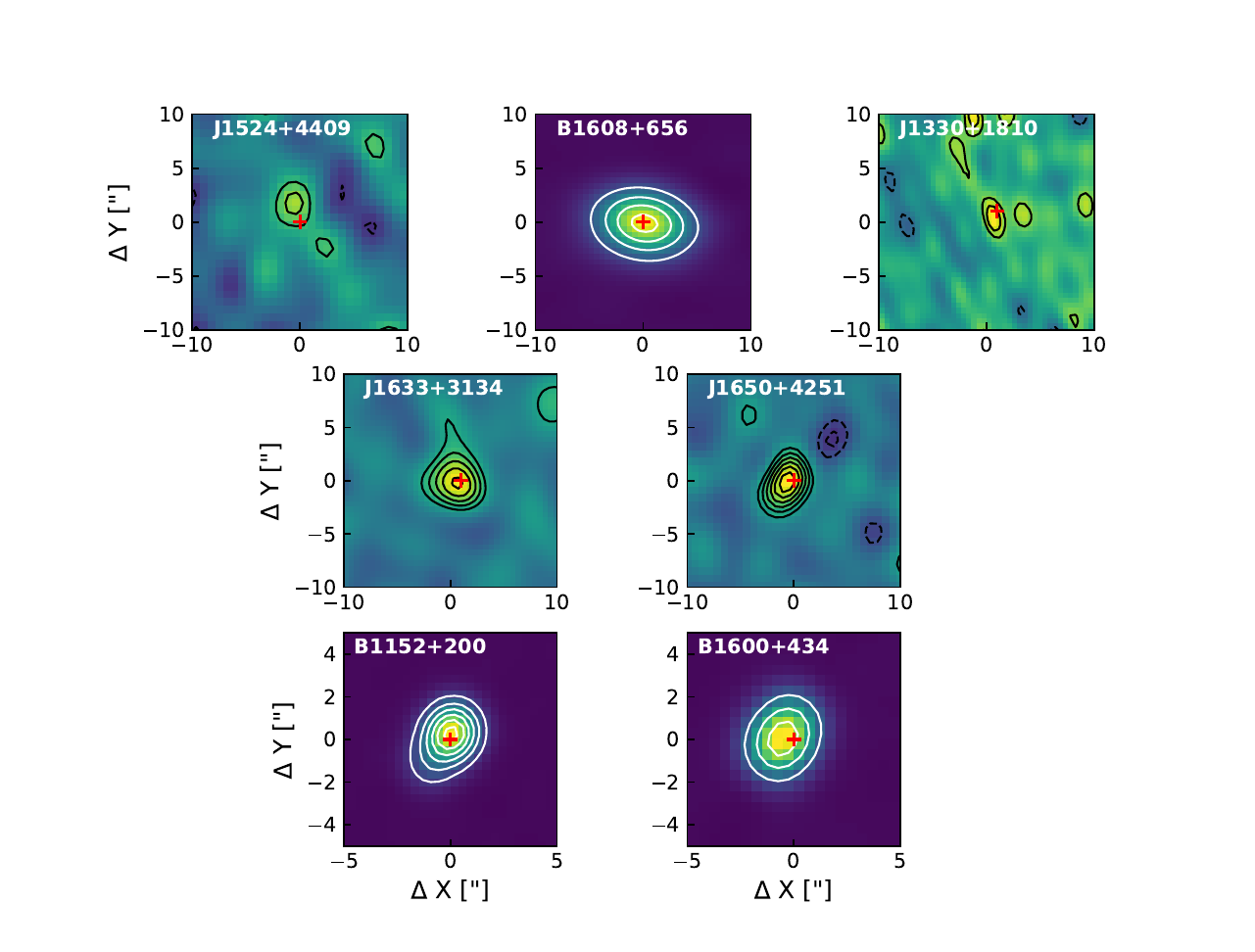}
    \caption{Detected continuum emission at the position of the CO(2--1) emission line for the sample presented in this work. Contours for J1524+4409, J1330+1810, J1633+3134, and J1650+4251 start at $\pm2\sigma$ and increase in intervals of $\pm1\sigma$, where $\sigma$ is given in Table \ref{tab:noema_beams}. Contours are given in steps of $\pm$50$\sigma$ starting at 50$\sigma$ for B1608 and in steps of $\pm$100$\sigma$ starting at 100$\sigma$ for B1152 and B1600. The red cross indicates the peak position of the gas emission where detected, and the optical emission for the three QSOs not detected in CO(2--1).}
    \label{detections_continuum}
\end{figure*}

\subsection{Rest-frame 1.3-mm continuum}\label{sec:cont}

We detected continuum emission in 5 of the 10 targets; two more targets showed tentative $\sim$2$\sigma$ emission and three were undetected (see Figure \ref{detections_continuum}). The continuum emission was unresolved, and this was confirmed by a curve of growth analysis which showed that the continuum emission was more compact than that of the CO line. Therefore, we extracted the flux, $S_\mathrm{cont}$, from an aperture with a diameter twice that of the beam FWHM, to ensure we recovered all the flux. The continuum fluxes are reported in Table \ref{tab:results}, corrected for magnification. We found a good agreement between $S_\mathrm{cont}$ and the flux density values estimated by fitting a two--dimensional Gaussian model to the continuum image (Fig. \ref{detections_continuum}). Two of our sources had already been observed at this same frequency during the survey carried out by \cite{barvainis2002-co} with the Plateau de Bure Interferometer, which gave us the opportunity to assess if the measured fluxes have varied, possibly induced by variable AGN activity. The flux densities quoted below were uncorrected for magnification.

B1608+656: this QSO at $z =$ 1.394 has a continuum flux of 18.4$\pm$0.7 mJy at 94~GHz and 16.0$\pm$0.7 mJy at 110~GHz, a factor of two larger than the value of 8.1$\pm$0.4 mJy reported by \cite{barvainis2002-co}.

B1600+434: this QSO at $z =$ 1.589 has a continuum flux of 34.3$\pm$0.1 mJy at 89~GHz and 31.1$\pm$0.1 mJy at 104~GHz, compared to the value of 25$\pm$0.3 mJy reported by \cite{barvainis2002-co}.

These two sources, together with B1152+200, are QSOs with strong radio emission \citep{Stacey2018}. It is thus likely that the 1.3mm continuum detected with NOEMA is also associated with synchrotron emission coming from the jets.

\begin{table*}[!htb]
\centering
\caption{Parameters derived for the sample of unobscured QSOs in this work.  Columns give the following information: bolometric luminosity, IR luminosity integrated between 8--1000$\mu$m and corrected for AGN contamination, stellar mass, IR--derived SFR,  gas mass, depletion time, gas fraction, fraction of the AGN contribution to the total IR luminosity, and 3$\sigma$ upper limits on mass outflow rates. The values were corrected for magnification as listed in Table \ref{tab:sample}. \label{tab:sed} }
\begin{tabular}{lccccccccc}
\hline 
\hline
Target & log($L_\mathrm{bol}$) & log($L_\mathrm{IR}$) & log(M$_*$) & SFR$_\mathrm{IR}$ & $M_\mathrm{gas}$ & $t_\mathrm{dep}$ & $f_\mathrm{gas}$ & $f_\mathrm{AGN}$ & $\dot{M}_\mathrm{H_2,out}$\\
 & [erg s$^{-1}$] & [L$_\mathrm{\odot}]$ & [M$_\mathrm{\odot}$] & [M$_\mathrm{\odot}$ yr$^{-1}$] & [M$_\mathrm{\odot}$($\times$10$^{9}$)] & [Myr] & & & $M_\odot$ yr$^{-1}$\\
\hline
J1524+4409 & 43.3$^{+0.9}_{-0.9}$ & 11.8$^{+0.2}_{-0.3}$ & 10.0$^{+0.2}_{-0.1}$& 94$^{+55}_{-44}$ & 30$\pm$1 & 290$^{+170}_{-130}$ & 0.73$^{+0.10}_{-0.04}$ & 0.07 & $<$660\\
B1608+656$^*$ & 42.8$^{+0.6}_{-0.6}$ & 11.2$^{+0.1}_{-0.1}$ & 11$^{+0.1}_{-0.1}$& 25$^{+5}_{-5}$ & $<$2 & $<$70 & $<$0.02 &  0.86 & --\\
J1330+1810 & 43.4$^{+0.6}_{-1.1}$ & 11.7$^{+0.1}_{-0.1}$& 9.5$^{+0.1}_{-0.2}$ & 80$^{+27}_{-14}$& 11$\pm$0.2 & 130$^{+50}_{-30}$ & 0.79$^{+0.07}_{-0.05}$ & 0.39 & $<$130\\
J1455+1447 & 44.5$^{+0.1}_{-0.8}$ & 11.7$^{+0.2}_{-0.1}$& 9.5$^{+1}_{-0.3}$& 70$^{+50}_{-18}$&  7$\pm$1.5 & 100$^{+80}_{-30}$ & 0.70$^{+1.9}_{-0.15}$ & 0.49 & $<$275\\
J1633+3134$^*$ & 46.1$^{+0.03}_{-0.1}$ & 12$^{+0.1}_{-0.2}$& 8.7$^{+1.1}_{-2.3}$& 160$^{+43}_{-63}$&  9$\pm$5 & 60$^{+30}_{-40}$ & 0.91$^{+0.55}_{-0.06}$ & 0.67 & $<$170\\
J0924+0219 & 44.0$^{+0.8}_{-1.4}$ & 11.6$^{+0.1}_{-0.1}$& 10$^{+0.2}_{-0.1}$& 61$^{+20}_{-18}$&  21$\pm$2 & 900$^{+300}_{-250}$ & 0.84$^{+0.08}_{-0.02}$ & 0.33 & $<$210 \\
J1650+4251 & 45.9$^{+0.1}_{-0.2}$ & 11.9$^{+0.1}_{-0.1}$& 10$^{+0.7}_{-0.2}$& 110$^{+40}_{-20}$&  15$\pm$8 & 130$^{+80}_{-70}$ & 0.52$^{+0.96}_{-0.08}$&  0.43 & $<$317\\
B1152+200$^*$ & 45.9$^{+0.1}_{-0.1}$ & 11.2$^{+0.4}_{-0.4}$ & 9.9$^{+0.5}_{-0.4}$& 26$^{+38}_{-16}$&  $<$2 & $<$70 & $<$0.20 & 0.83 & --\\
B1600+434$^*$ & 44.7$^{+0.1}_{-0.2}$ & 11.4$^{+0.1}_{-0.1}$ & 10.4$^{+0.3}_{-0.4}$ & 34$^{+5}_{-6}$& $<$2 & $<$60 & $<$0.06 & 0.07 & --\\
J0806+2006 & 43.0$^{+0.7}_{-1.5}$ & 11.8$^{+0.2}_{-0.3}$& 10.2$^{+0.3}_{-0.1}$& 88$^{+43}_{-48}$& 4$\pm$2 & 50$^{+30}_{-40}$ & 0.20$^{+0.15}_{-0.06}$ &0.30 & $<$450\\
\hline
\multicolumn{10}{l}{$^*$Sources are classified as radio--loud QSOs \citep{Stacey2018}}\\
\hline 
\end{tabular}
\end{table*}

\subsection{Upper limits on outflow flux}

We inspected the continuum--subtracted data cubes to search for high--velocity line emission that might be originating from outflows in the sources with detected CO(2--1) emission. We did not find any significant ($>$3$\sigma$) emission in any of the high--velocity channels that did not contain emission from the main galaxy (beyond $\pm$400 km s$^{-1}$). To place an upper limit on the total flux, we created intensity maps by collapsing the velocity channels from +500 km s$^{-1}$ to +1000 km s$^{-1}$ and from -500 km s$^{-1}$ to -1000 km s$^{-1}$, a range commonly adopted when searching for outflow emission in previous works \citep{cicone2014,lutz2020,bischetti2019a}. We used the same circular aperture of 6$"$ radius centred on the CO(2--1) emission to extract the flux. This ensured that we were at least covering the area where CO emission was detected, placing conservative upper limits on the mass outflow rates (see Section \ref{sec:Lco}). We did not detect $>$2$\sigma$ emission in any of the targets. We note that, using the AIC criterion, a double Gaussian was preferred to fit the spectrum of J1524+4409, in a configuration that is often indicative of an outflow. However, it is not unusual to find asymmetric line profiles in lensed sources due to differential magnification of different velocity components \citep{butler2023}, so we cannot currently confirm the presence of an outflow in this source.

\subsection{Spectral energy distribution decomposition}\label{sec:sed}

To derive the intrinsic properties of the QSO host galaxies, we used multi-wavelength spectral energy distribution (SED) modelling. Specifically, we included ultraviolet to mm-wave photometry\footnote{We excluded the X-ray data due to the QSO variability, and we also excluded the radio data as they are not included in \textsc{AGNfitter}.} available in the literature and via the NASA NED database. As the listed uncertainties are likely to be underestimated due to the use of different apertures, flux calibration uncertainties or confusion limits, for example, we added a 10\% error in quadrature to the photometry.

The main challenge in SED modelling of these multi-wavelength datasets is that at most wavelengths, the low spatial resolution of the data causes the blending of the foreground (lensing) galaxy (S$_\mathrm{FG}(\lambda)$) with the background AGN (S$_\mathrm{AGN}(\lambda)$) and host galaxy itself (S$_\mathrm{BG}(\lambda)$):

\begin{equation}
S_\mathrm{obs}(\lambda) = S_\mathrm{FG}(\lambda) + \mu (S_\mathrm{AGN}(\lambda) + S_\mathrm{BG}(\lambda)).
\end{equation}

To decompose these three components, we use a three-step approach:
\begin{itemize}
    \item We used resolved imaging (mostly at optical/near--IR wavelengths) to separate the light from the foreground galaxy.
    \item We fit the light from the foreground galaxy using the \citep{brown2014} templates of nearby galaxies (spanning rest-frame UV to MIR wavelength for a broad range of galaxy types, including ellipticals, spirals, merging galaxies, blue compact dwarfs, and luminous infrared galaxies) and subtracted these from the total SED. This removed the $S_\mathrm{FG}$ component from the data. We note that the templates used did not extend out to MIR and FIR data. All the lensing galaxies in our sample, with the exception of the lens of B1600+434, are ellipticals with little or no star formation left, and so we did not expect them to have a significant contribution to the total FIR emission. For B1600+434, we used the available SPIRE photometry for UGC 12150, NGC 5104, NGC 5033, and NGC 4594, four of the best fitting spiral templates from the \citep{brown2014} catalogue, to extend the average subtracted model and account for the extra FIR emission.
    \item We model the remaining $\mu$($S_\mathrm{AGN}(\lambda) + S_\mathrm{BG}(\lambda)$) SED using \textsc{AGNFitter}, a publicly available SED-modelling package specifically designed to model the SEDs of galaxies with prominent AGNs \cite{Calistro2016}. The photometry is fitted with a combination of a stellar component, an optical and an infrared AGN component and a FIR component associated with dust-obscured star formation in the host galaxy \footnote{Independent \textsc{AGNFitter} models for the spectral energy distributions of J0924+0219 and J1330+1810 were recently presented by \cite{stacey2022}. Compared to the Stacey et al. models which assumed that the optical emission is dominated by the QSO component, we removed the foreground contamination and used wider priors which allow optical emissions to be dominated by the stellar component. On average, our approach yielded slightly lower QSO luminosities in the optical.}. We used the spectroscopic redshifts derived from fitting the CO(2--1) emission lines as input where available. We refer the reader to \cite{Calistro2016} for a detailed description of the algorithm and templates used.
\end{itemize}

We derived bolometric and infrared luminosities, stellar masses and IR-based SFRs from SED fitting. The results are summarised in Table \ref{tab:sed}, and the SED fits are shown in Appendix \ref{appendix:sed_fits_app}. Following \cite{calistro2021}, we calculated bolometric luminosities $L_\mathrm{bol}$ by integrating the emission from the accretion disc component (BBB) over the wavelength range 0.05−-1$\mu$m, with an added correction factor of $\Delta$log($L_\mathrm{bol}$) = 0.3 to account for X-ray emission. The inferred $L_\mathrm{bol}$ are consistent with those derived from the 3000\AA \ luminosity \citep{rakshit2020}\footnote{For all QSOs except B1608+656 and B1600+434, which were not in the catalogue in \cite{rakshit2020}.}. The infrared luminosity, $L_\mathrm{IR}$, was calculated by integrating the galaxy cold dust emission over the wavelength range 8--1000$\mu$m after subtracting the AGN contribution to the emission \citep[e.g.][]{mullaney2011,delmoro2013,delvecchio2014}. The infrared luminosities obtained are comparable to those derived by \cite{Stacey2018} (after correcting for $L_\mathrm{IR} = 1.91L_\mathrm{FIR}$) for the QSOs with CO detections, with a median ratio of 1.0$\pm$0.2 between both values. The QSOs not detected in CO show a larger discrepancy, with the \textsc{AGNfitter}--derived infrared luminosities being on average $\sim$40\% lower than calculated by \cite{Stacey2018}. We note that the available photometry for the QSOs does not currently sample the peak of the IR SED. This is evidenced by the large range of possible luminosities covered by the \textsc{AGNfitter} models (green component, Figs. \ref{fig:sed1}, \ref{fig:sed2}).

In order to derive the SFR from the starburst model component dominating the FIR, we follow \cite{murphy2011}:
\begin{equation}
    \mathrm{SFR}\ [\mathrm{M_\odot \ yr^{-1}}] = 10^{-10} \times \mathrm{L_\mathrm{IR} \ [L_\odot]},
\end{equation}

\noindent corrected to a Chabrier initial mass function \citep{chabrier2003}. This does not account for unobscured star formation, so it should be considered as a lower limit. The large uncertainties in the stellar masses reflect the difficulty in separating the optical emission from the stellar population and the QSO in the lensed galaxy. To test the robustness of the stellar masses, we re--ran \textsc{AGNfitter} without deblending the emission from the foreground and background galaxies (that is, we assumed that all light was associated with the background galaxy). We found that the stellar masses are consistent within a factor of 2. Finally, we used the BH masses (M$_\mathrm{BH}$) derived from SDSS spectroscopic observations of the MgII emission line \citep{rakshit2020} and the $M_\mathrm{BH}-M_*$ scaling relation from \cite{ding2020} to calculate the expected values for the stellar mass. The median ratio between the SED- and M$_\mathrm{BH}$-derived stellar masses is 1.05, with a mean ratio of 3.8.

\subsection{Spectral stacking - Upper limits on dense-gas tracers} \label{sec:hcn}

\begin{figure}
    \centering
    \includegraphics[width = 0.49\textwidth]{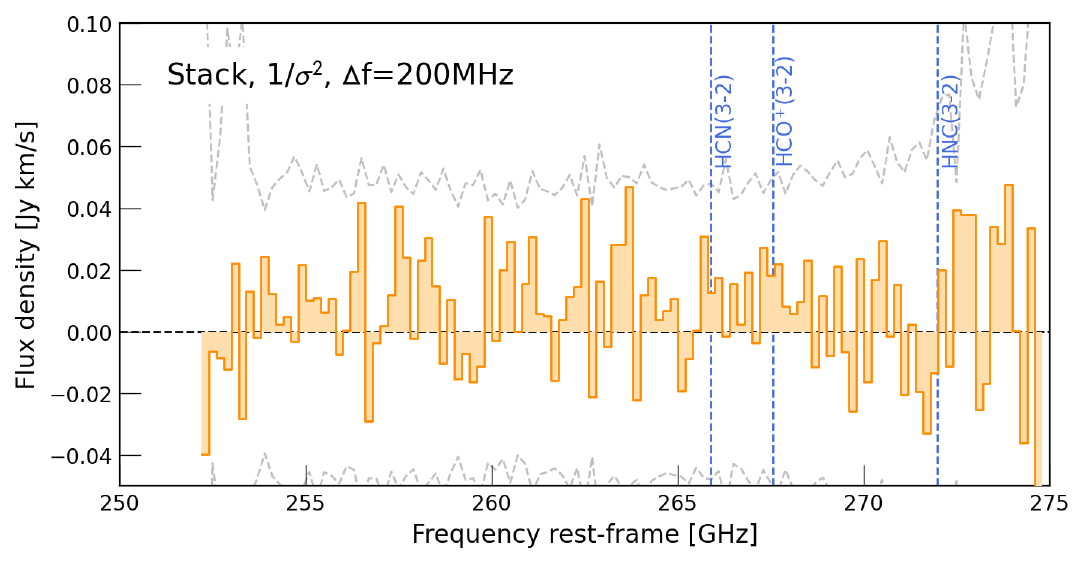}
    \caption{Stacked spectrum of our NOEMA observations over the region covering the HCN/HCO$^+$/HNC(3--2) lines. The spectrum was scaled to an apparent FIR luminosity $L_\mathrm{FIR}=5\times10^{12}$~$L_\odot$. We did not find any line detected at $\geq$3$\sigma$ significance, indicated by the dashed horizontal line.}
    \label{stacked_spectrum}
\end{figure}

Our spectral setup covers several other emission lines, notably HCN(3--2), HCO$^+$(3--2) and HNC(3--2). As none of the lines was detected in individual spectra, we resorted to spectral stacking, following the same procedure as in \citet{Rybak2022}: (1): first, we shifted the spectra in frequency to a common redshift $z=1.25$; (2) we scaled the observed flux densities to match the luminosity distance at $z=1.25$; (3) we took a weighted mean using $1/\sigma_\mathrm{rms}^2$ weighting. Fig.~\ref{stacked_spectrum} shows the resulting rest-frame stacked spectrum, normalised to a $L_\mathrm{FIR}=5\times10^{12}$~$L_\odot$. We did not see any excess line emission, independent of the choice of weighting and spectral binning.

Based on the stacked spectrum, we inferred the following $3\sigma$ upper limits on the line vs FIR luminosity ratios: $L'_\mathrm{HCN(3-2)}/L_\mathrm{FIR} \leq 2\times10^{-4}$ K km/s pc$^2$/$L_\odot$, $L'_\mathrm{HCO^+(3-2)}/L_\mathrm{FIR}2\times10^{-4}$ K km/s pc$^2$/$L_\odot$, $L'_\mathrm{HNC(3-2)}/L_\mathrm{FIR}\leq 2.6\times10^{-4}$ K km/s pc$^2$/$L_\odot$. These are consistent with the ratios observed in local (ultra) luminous infrared galaxies (e.g. \citealt{Li2020}).

\section{Analysis} \label{sec:Analysis}

To investigate the physical properties of our sample, we compile the following QSOs with observations of molecular gas from the literature for comparison: Q 0957+561 and HS 0810+2554, the only $z\sim$1.5 unobscured QSOs with CO line emission detections in the literature \citep{krips2005,chartas2020}; 34 unobscured QSOs at $z = 1.3--6.7$ and 47 obscured QSOs at $z = 1.1--6.4$ from the compilation by \cite{perna2018}; seven X-ray selected QSOs at $z\sim$2 drawn from SUPER (SINFONI Survey for Unveiling the Physics and Effect of Radiative feedback), which have a reliable measure of $L_\mathrm{IR}$ \citep{circosta2021}; seven unobscured QSOs (bolometric luminosity $L_\mathrm{bol}$ $>$ 3$\times$10$^{47}$ erg s$^{-2}$) from the WISE-SDSS selected hyper-luminous (WISSH) QSOs sample at $z\sim$2.4--4.7 from \cite{bischetti2021}; 23 $z<$0.1 Palomar--Green QSOS from \cite{shangguan2020}. Finally, we include a sample of 17 SMGs from the AS2COSMOS, AS2UDS and AEGIS surveys presented by \cite{friascastillo2023} and 16 SFGs at 1$<z<$3 detected in CO emission with the ASPECS programme \citep{boogaard2020}. The stellar masses and SFRs for these samples were uniformly obtained through SED fitting with \textsc{magphys} \citep{dacunha2008,dacunha2015}.

Where necessary, the reported quantities have been corrected for magnification. In order to homogenise the data, we correct the gas masses associated with the QSOs for $\alpha_\mathrm{CO}$=4. For the SMGs, the most recent studies suggest $\alpha_\mathrm{CO}$=1 is more appropriate given the physical properties of their ISM \citep{birkin2020}, so we calculate their corresponding total gas masses according to this value. For a more detailed discussion on the adopted $\alpha_\mathrm{CO}$, see Section \ref{sec:Lco}.

The unobscured QSOs targeted by this work were originally drawn from a variety of surveys at optical and radio wavelengths, making it hard to identify and quantify the selection effects biasing the derived properties. However, although small, our sample comprises half of all the lensed QSOs in the $z$ = 1 -- 1.5 range from the parent sample in \cite{Stacey2018}, with the other half being \textit{Herschel}--faint sources with no indication of significant ongoing star formation. Therefore, it is likely that our sources are representative of the star--forming unobscured QSO population at this epoch.

\subsection{Unobscured QSOs on the SFR--M$_*$ plane} \label{main-sequence}

Normal star--forming galaxies have been shown to follow a tight correlation between their star formation rates and stellar masses, known as the `star--forming main--sequence (MS)', which evolves with redshift \citep[e.g.][]{noeske2007,elbaz2011,speagle2014,schreiber2015,leslie2020}. Here, we consider where our targets and the comparison samples lie in relation to the MS as parameterised by \cite{speagle2014} at $z=1.25$. The positions of all the literature samples have been scaled to a common redshift of $z=1.25$. Since the MS evolves with redshift, this scaling is necessary to avoid higher--redshift sources appearing like starbursts when placed on the lower--redshift MS relation. It is done by plotting each source on the $z=1.25$ MS at the same $\Delta$MS that it would have at its true redshift, where $\Delta$MS is the ratio of its measured SFR compared to that of the main sequence at its redshift and stellar mass, $\Delta$MS = SFR/SFR$_\mathrm{MS(z,M_*)}$. Following the literature \citep[e.g.][]{rodighiero2011}, we classify a galaxy as a starburst if it lies $\Delta$MS$>$0.6 dex above the main sequence.

\begin{figure} [!htb]
    \centering
    \includegraphics[scale=0.45]{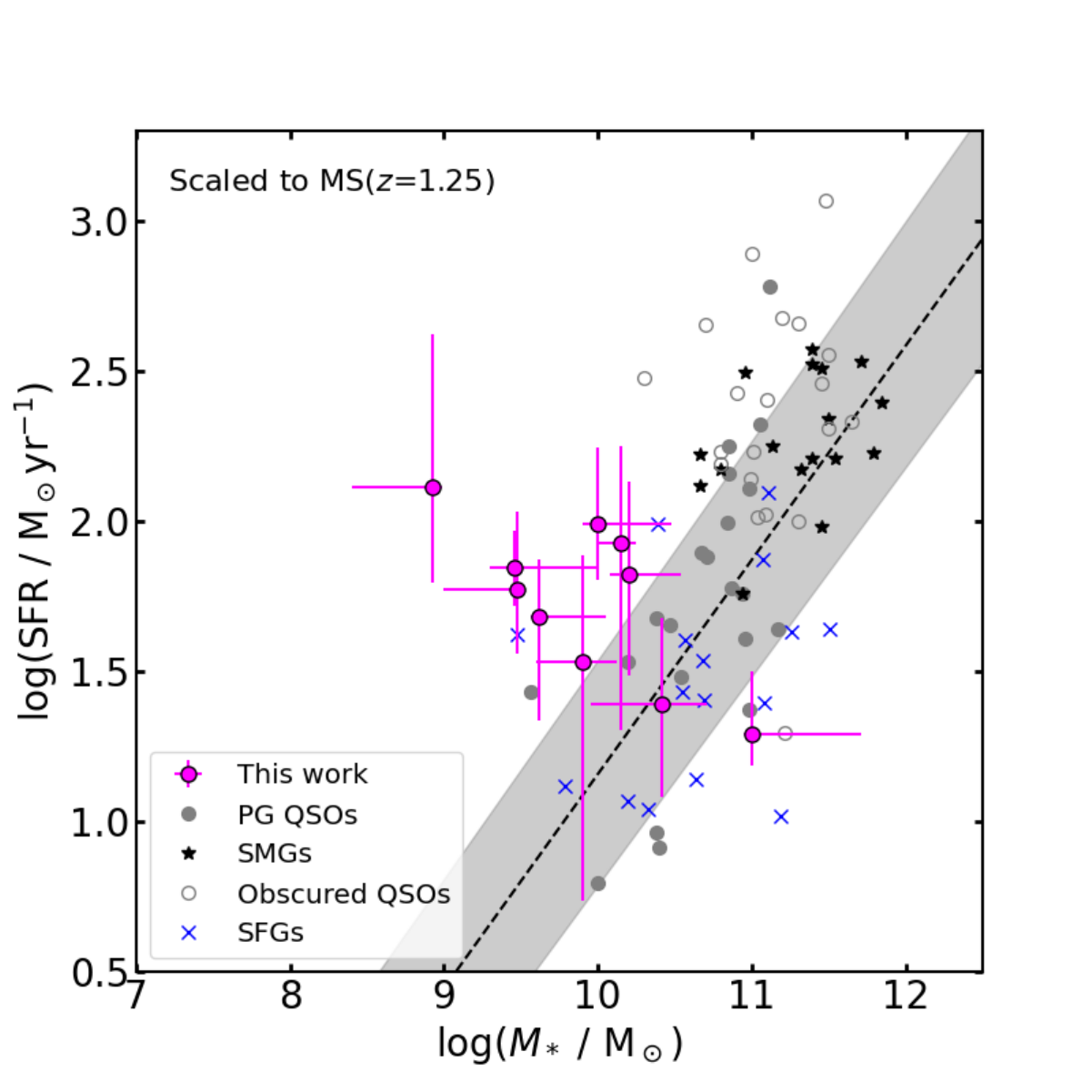}
    \caption{SFR as a function of stellar mass for our sample of unobscured QSOs. The values were obtained through SED fitting with \textsc{AGNfitter}. For comparison, we have plotted the low--redshift Palomar--Green QSOs from \cite{shangguan2020} (grey), high--redshift obscured QSOs (grey, empty circles) from the compilation presented in \cite{perna2018}, a sample of high--redshift SMGs (black stars) and $z=$1--3 SFGs from ASPECS \citep[blue crosses,][]{boogaard2020}. The dashed black line marks the main--sequence at $z=$1.5 as parameterised by \cite{speagle2014}, with the one sigma scatter shown by the shaded grey area. All data points have been scaled to $z=1.25$ for comparison and corrected for magnification where necessary. Most of our targets lie on or above the main--sequence. Their stellar masses are lower than most of the higher--redshift obscured QSOs, but comparable to low--redshift unobscured QSOs and the ASPECS SFGs.}
    \label{main_sequence}
\end{figure}

We show the distribution of our targets on the SFR--$M_*$ plot in Fig. \ref{main_sequence}. Six of our targets have high SFRs and are classified as starbursts following the above definition. The IR--based SFRs are in the range 25--160 $M_\odot$ yr$^{-1}$, with a median of $\sim$75 $M_\odot$ yr$^{-1}$. For comparison, \cite{stanley2017} and \cite{symeonidis2022} find IR--based SFRs of 25--120 and $\sim$60 $M_\odot$ yr$^{-1}$ for $0.8<z<1.5$ and $z=1.5$ optically selected QSOs, respectively. The SFRs are also consistent with those of the ASPECS SFGs at the same redshift. Therefore, our QSOs do not appear to be exceptional in terms of SFRs compared to other systems at similar redshifts. These results support the consensus that star formation can co--exist with AGN activity, and any significant suppression of star formation due to negative feedback effects must occur on longer timescales \citep{floyd2013,Stacey2018,rodighiero2019,carraro2020,jarvis2020}.

The stellar masses, with a mean of M$_* = 2\times10^9 M_\odot$, are consistent with the ASPECS SFGs and low--redshift PG unobscured QSOs, but are on average an order of magnitude lower than those of high--redshift obscured QSOs and SMGs. This could be due to gravitational lensing bias -- since the SMGs and obscured QSOs are not lensed, we are biased towards galaxies that are massive and bright enough to be above the detection limits of the different studies. Gravitational lensing allows us to pick up less massive systems that would otherwise be below the detection threshold. We find specific SFRs (sSFR = SFR / $M_*$) in the range log(sSFR) = --6.7 to --9.6, with a median log(sSFR) of --8.1.

\subsection{$L'_\mathrm{CO}$ versus $L_\mathrm{IR}$} \label{sec:Lir}

\begin{figure*}
    \centering\includegraphics[scale=0.45]{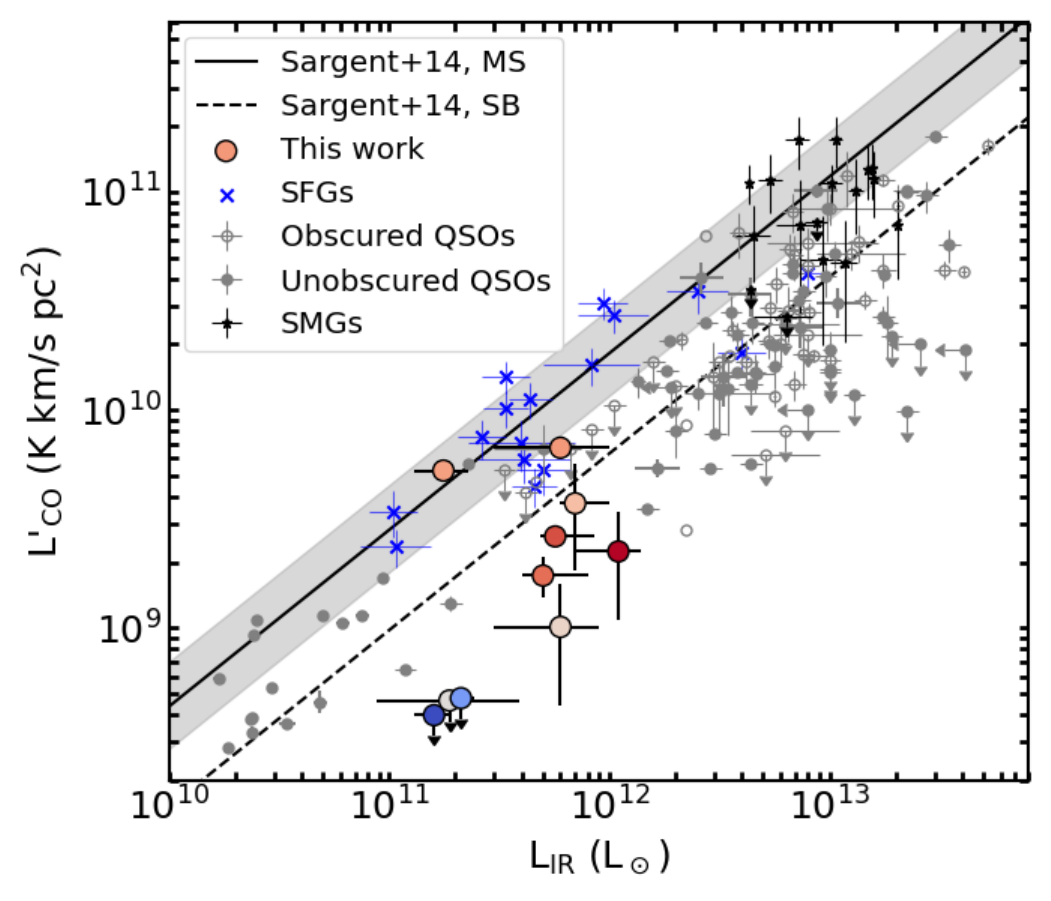}
    \includegraphics[scale=0.45]{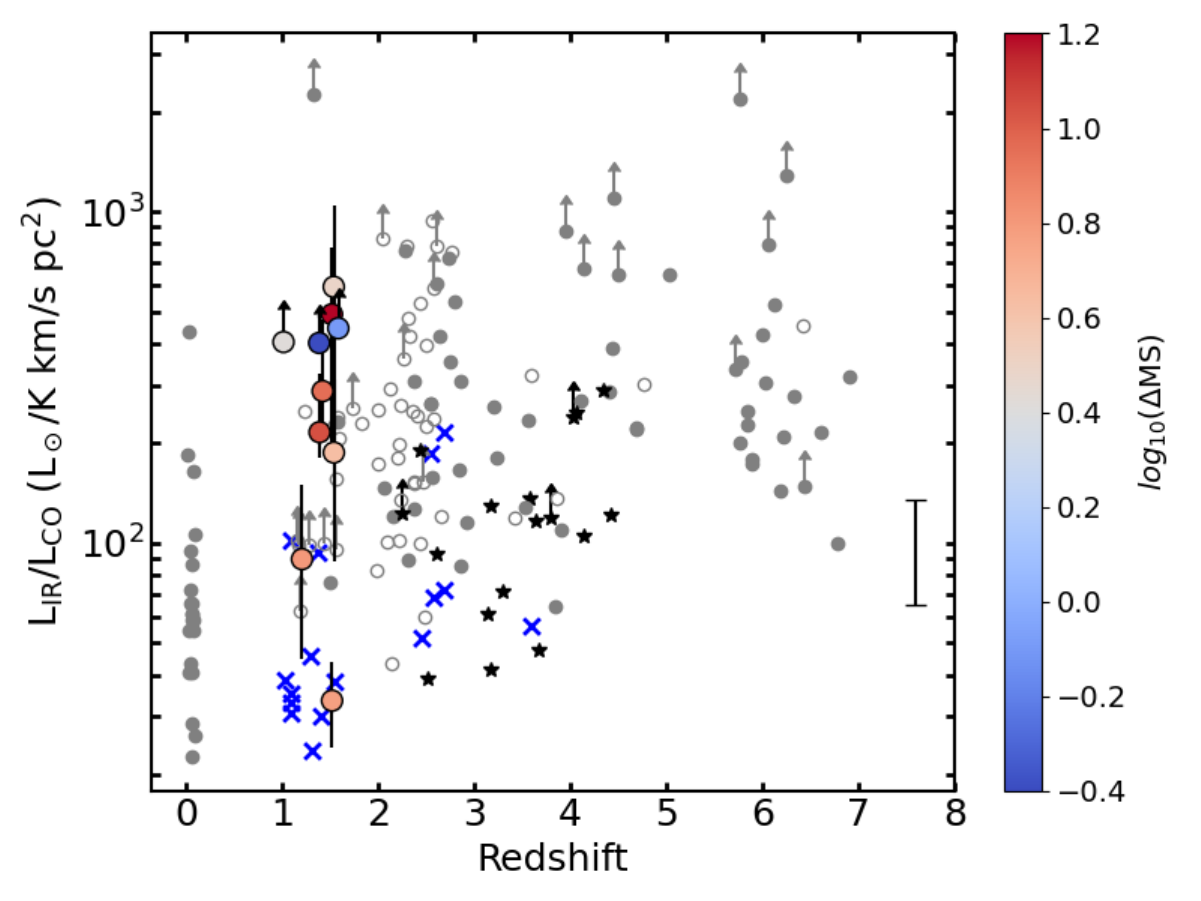}
    \caption{Left: CO(1--0) line luminosity versus infrared luminosity integrated over 8--1000$\mu$m for our sample of unobscured QSOs, coloured as a function of offset from the MS. The CO(1--0) luminosities have been calculated using $r_\mathrm{21}=0.86$ \citep{Kirkpatrick2019}, and $L_\mathrm{IR}$ has been corrected for AGN contamination. For comparison, we have plotted the low--redshift Palomar--Green QSOs from \cite{shangguan2020}, high--redshift obscured (filled grey circles) and unobscured (open grey circles) QSOs from the compilation presented in \cite{perna2018}, a sample of high--redshift SMGs (black stars) and $z=$1--3 SFGs from ASPECS \citep[][blue crosses]{boogaard2020}. Where necessary, all quantities have been corrected for magnification. The black solid and dashed lines show the $L'_\mathrm{CO}--L'_\mathrm{IR}$ relation expected for main sequence and starburst galaxies at $0<z<3$ from \cite{sargent2014}, respectively. Similar to other samples of QSOs at both low and high redshift, most of the $z\sim1.5$ QSOs lie below the relation for main sequence galaxies. Right: Ratio of $L_\mathrm{IR}$/$L'_\mathrm{CO}$, used as a probe for star formation efficiency, versus redshift. The vertical line on the right shows the mean error of the comparison samples. Our targets show a large scatter, with a median $L_\mathrm{IR}$/$L'_\mathrm{CO}$ = 350$\pm$170 L$_\odot\ \mathrm{K \ km \  s^{-1} \ pc^2}$. For comparison, high--redshift QSOs have a ratio of $L_\mathrm{IR}$/$L'_\mathrm{CO}$ $\sim$250, and SMGs show $L_\mathrm{IR}$/$L'_\mathrm{CO}$ $\sim$100.}
    \label{sfe}
\end{figure*}

We derive CO(2--1) line luminosities following \cite{solomon2005}: 
\begin{equation}
    L\mathrm{'{_{CO}}} = 3.25 \times 10^7 \ I\mathrm{{_{CO}}} \ \mathrm{\nu{^{-2}_{obs}}} \ D\mathrm{{^2_{L}}} \ (1+\mathrm{z})^{-3} \ \mathrm{K \ km \  s^{-1} \ pc^2} ,
\end{equation}

\noindent where $I_\mathrm{\text{CO}}$ is the integrated line flux from the 0th-moment map in Jy km s$^{-1}$, $\mathrm{\nu_{\text{obs}}}$ is the observed frequency in GHz and $D\mathrm{_{\text{L}}}$ is the luminosity distance in Mpc. In order to derive the CO(1--0) line luminosities, we assume an excitation correction factor, r$_\mathrm{21}$ = $L'_\mathrm{CO(2-1)}$/$L'_\mathrm{CO(1-0)}$ = 0.86 \citep{Kirkpatrick2019}.

Fig. \ref{sfe} (left) shows the integrated CO luminosities of our targets and the comparison samples as a function of $L_\mathrm{IR}$, where the latter has been corrected for AGN contamination. We show the best--fit relation derived for main--sequence galaxies at 0$<z<$3 by \cite{sargent2014}. Both luminosities have been shown to correlate strongly at both high and low redshift \citep{sanders_mirabel1985,solomon2005,genzel2010,saintonge2011,carilli-walter2013}. Previous studies at $z\geq$2 have found that, while both luminosities are also correlated for QSOs, these appear to be deficient in CO luminosity for a given IR luminosity compared to normal, star--forming galaxies \citep[e.g.][but see \citealt{Kirkpatrick2019}]{perna2018,bischetti2021,circosta2021}. Our QSOs show intermediate $L'_\mathrm{CO}$ and $L_\mathrm{IR}$ compared to their high-- and low--redshift counterparts. J1524+4409 and J0924+0219 are consistent with the \cite{sargent2014} relation for star--forming galaxies, while five of them have lower $L'_\mathrm{CO}$ values than the relation would predict for their $L_\mathrm{IR}$, in line with other high--redshift QSO studies. It is possible that the QSOs follow instead the relation found for starbursts, rather than main--sequence galaxies \citep[approximately a factor of 3 below the locus of MS galaxies][]{sargent2014}. J1650+4251 is consistent with the relation for starbursts, but the remaining four detections and three non--detections still fall below the relation. This deficit in CO luminosity compared to IR luminosity has been suggested as a possible sign of AGN feedback, which would shut off star formation by either depleting the gas present in the host galaxy \citep{perna2018,bischetti2021} and/or heating up the cold gas through the injection of energy and momentum via outflows.

The three non--detected QSOs also have a large offset from the $L'_\mathrm{CO}$--$L_\mathrm{IR}$ relation, as well as the lowest offset from the main--sequence (B1152+656 can be classified as quiescent). This combination of low gas mass and star formation rate could be explained if the host galaxies are at the end of the blowout phase, where the molecular gas has been depleted by either QSO or star formation feedback and they are now entering the quenched, post--starburst phase. Theoretical models and observations of nearby galaxies have also shown X-ray irradiation from the central AGN to dissociate CO molecules \citep[e.g.][]{malloney1996,izumi2020}, enhancing the abundance of carbon atoms and further contributing to the CO deficit, but observations of atomic carbon or CO isotopologues would be needed to explore this scenario.

As the IR luminosity is a tracer for star formation, the ratio between $L'_\mathrm{CO}$ and $L_\mathrm{IR}$ serves as a proxy for star formation efficiency, usually defined as SFE = SFR/M$_\mathrm{gas}$ in units of yr$^{-1}$. We show this ratio for our sample in Fig. \ref{sfe} (right). Our QSOs show SFE ratios in the range 30--560 L$_\mathrm{\odot}$/(K km s$^{-1}$ pc$^2$), with a median of 350 L$_\mathrm{\odot}$/(K km s$^{-1}$ pc$^2$). This is similar to the SFE of most high--redshift QSOs and SMGs, but elevated compared to the ASPECS SFGs (median $\sim$60 L$_\mathrm{\odot}$/(K km s$^{-1}$ pc$^2$)) or the low--$z$ PG QSOs (median $\sim$50 L$_\mathrm{\odot}$/(K km s$^{-1}$ pc$^2$)) . There is likely not a single cause for the high SFE, as it is governed by the balance between the warm and cold HI phases, H$_2$ formation, and perhaps shocks and turbulent fluctuations driven by stellar and AGN feedback.  If these QSOs are indeed the last stage of a merger between two gas—rich galaxies, as galaxy evolution models predict, there are also extra factors that have been shown to affect the efficiency of star formation, such as the strength of the torques during the later stages of the merger or the geometry of the collision \citep{dimatteo2007,somerville2008}. It is also possible that there is an increased amount of dense gas available in the QSOs, as a consequence of either the merger or AGN and stellar feedback.  Higher-resolution, multi-wavelength data would allow us to better constrain the physical conditions driving the high SFE in these unobscured QSOs.

\subsection{$L'_\mathrm{CO}$ and gas masses} \label{sec:Lco}

In order to calculate the CO(1--0) line luminosities and derive total cold molecular gas masses from them, it is necessary to assume an excitation correction factor, $r_\mathrm{21}$ = $L'_\mathrm{CO(2-1)}$/$L'_\mathrm{CO(1-0)}$, and a conversion factor, $\alpha_\mathrm{CO}$ = $M_\mathrm{gas}$/$L'_\mathrm{CO(1-0)}$. For the excitation correction, we assume $r_\mathrm{21}$ = 0.86 \citep{Kirkpatrick2019}. The conversion factor $\alpha_\mathrm{CO}$ introduces the largest uncertainty when calculating gas masses, as it has been shown to be dependent on physical parameters such as metallicity, cloud density and temperature \citep{bolatto2013,acurso2017}. When there are no estimates of these parameters, however, the standard approach is to use either a Milky Way--like value of $\sim$4 M$_\odot$/(K km s$^{-1}$ pc$^{2}$) for normal star--forming galaxies, or a starburst--like factor of 0.8 M$_\odot$/(K km s$^{-1}$ pc$^{2}$) for highly star--forming or interacting galaxies \citep[e.g.][]{brusa2018,perna2018}. 

The standard approach in previous studies of molecular gas in QSOs has been to assume $\alpha_\mathrm{CO}$ = 0.8, justified by the fact that many have been found to have compact disc sizes \citep{brusa2018,feruglio2018}, high molecular gas excitation \citep[e.g.][]{wang2016} and reside in highly star-forming host galaxies \citep{carilli-walter2013,combes2018}. However, this might be biased by the fact that we have so far been limited to the most luminous QSOs at high redshift, and conditions affecting the value of $\alpha_\mathrm{CO}$ might evolve at lower redshifts. Indeed, \cite{Paraficz2018} derived $\alpha_\mathrm{CO}$ = 5.5$\pm$2.0 from resolved CO(2--1) observations for the $z=$0.65 QSO RXJ1131--1231, and \cite{shangguan2020} found that $\alpha_\mathrm{CO} \sim$ 3 is more appropriate for the low--redshift unobscured Palomar Green QSOs. This suggests that $\alpha_\mathrm{CO}$ = 0.8 might be too low a value to use for our sample. \cite{dunne2022} recently found $\alpha_\mathrm{CO}$ = 4 M$_\odot$ (K km s$^{-1}$ pc$^{2}$)$^{-1}$ based on the analysis of a sample of 407 galaxies ranging from local galaxies to high--redshift SMGs spanning up to $z \sim$ 6, so we assume this value for the purposes of the discussion. If we chose $\alpha_\mathrm{CO}$ = 0.8 instead and a steeper CO SLED \citep[e.g. $r_{21}$=0.99][]{carilli-walter2013}, the derived gas masses would be reduced by a factor of $\sim$6. In order to perform a uniform comparison and minimise systematic uncertainties, we calculate molecular gas masses using the same $\alpha_\mathrm{CO}$ = 4 for all QSOs from the literature samples.

\begin{figure}[htb!]
    \centering
    \includegraphics[width=\linewidth]{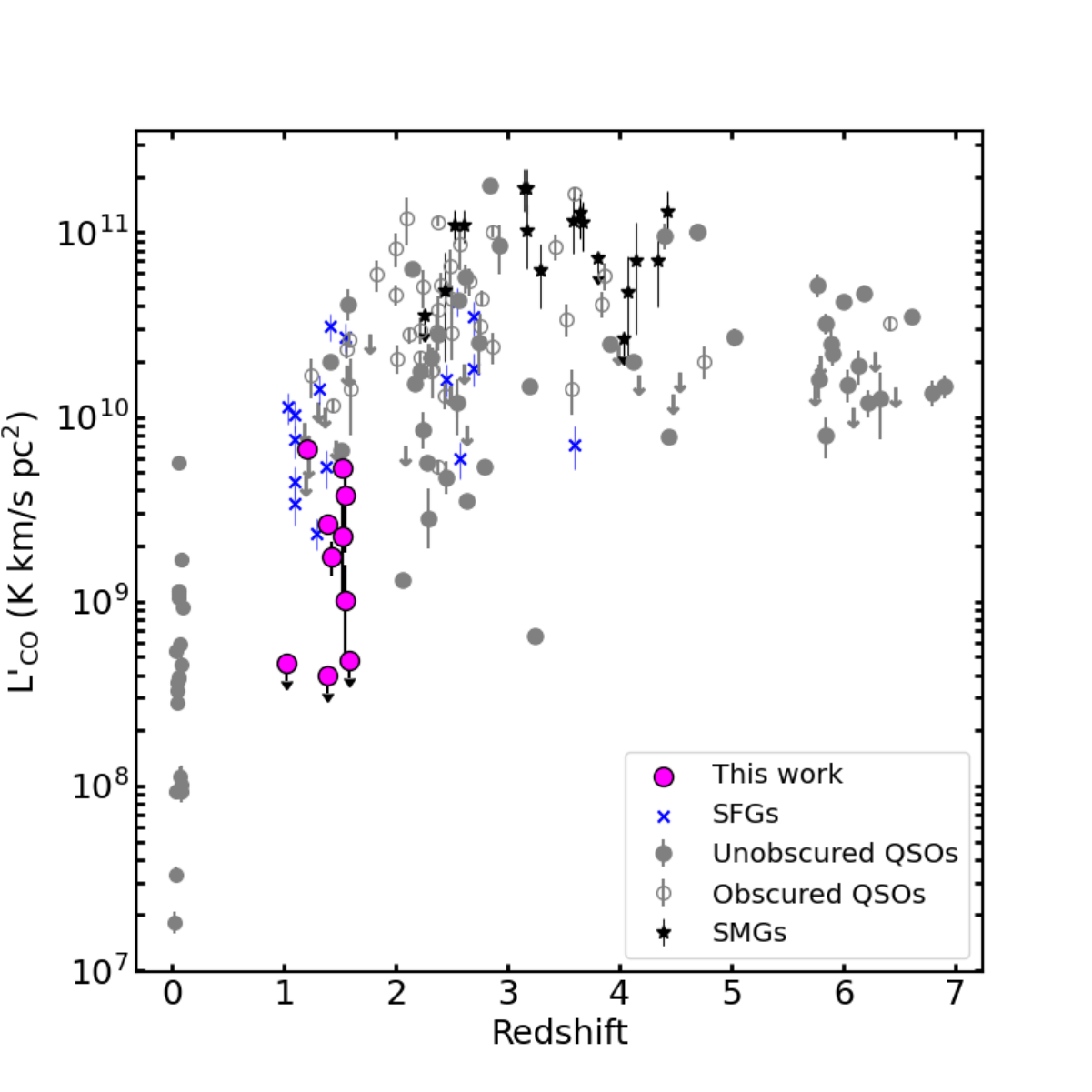}
    \caption{CO(1--0) line luminosities as a function of redshift for our sample. In some cases, error bars are smaller than the marker. The comparison samples are described in Fig. \ref{sfe}. Where necessary, all quantities have been corrected for magnification. On average, the unobscured QSOs targeted in this study have lower CO luminosities than other obscured QSOs and star--forming galaxies at similar redshift. It is thanks to gravitational lensing that we are able to detect these systems with short exposure times.}
    \label{lco_v_z}
\end{figure}

The CO(1--0) luminosities of our sample are shown as a function of redshift in Fig. \ref{lco_v_z}. We obtain line luminosities in the range $L'_\mathrm{CO(1-0)} \leq$0.4--6.7 $\times$ 10$^{9}$ $\mathrm{K \ km \  s^{-1} \ pc^2}$, corrected for magnification, with a median of $L'_\mathrm{CO(1-0)} =$ 2.0 $\times$ 10$^{9}$ $\mathrm{K \ km \  s^{-1} \ pc^2}$, corresponding to total cold molecular gas masses in the range $M_\mathrm{gas}\leq2$--40 $\times$ 10$^{9}$ M$_{\odot}$. Our sample of unobscured QSOs has therefore lower CO luminosities, and gas masses, than those of main--sequence galaxies and other obscured QSOs at similar redshift, but they are comparable to the most luminous QSOs in the local universe (for example, Palomar Green QSOs, \citealt{shangguan2020}). It is possible that, following the canonical galaxy evolutionary models, our unobscured QSOs are in a later evolutionary stage than obscured QSOs at similar redshifts, shown by their more depleted gas reservoirs. However, since the stellar masses are also lower than those of obscured QSOs (Fig. \ref{main_sequence}), we could also be probing a younger and/or less massive population of QSOs that is not evolutionary connected to the more luminous, high-$z$ counterparts studied so far. This could be an example of `galaxy down--sizing' \citep{cowie1996,fontanot2009,mortlock2011}, where more massive galaxies form their stars earlier and over a shorter period of time than lower--mass galaxies. Naturally, if the massive, high--redshift QSOs have exhausted their molecular gas reservoirs and quenched their star formation by $z\sim$1.5, they would have fallen below the detection threshold of the \textit{Herschel} survey \citep{Stacey2018}.

Finally, we can place conservative upper limits on the mass outflow rates of these QSOs based on the measured flux at $\pm$500 -- 1000 km s$^{-1}$. Assuming a simple spherical geometry uniformly populated with the outflowing clouds, the mass outflow rate can be calculated using \citep{maiolino2012,cicone2014}:
\begin{equation}
    \dot{M}_\mathrm{H_2,out} = 3v\frac{M_{H_2,out}}{R_{out}},
\end{equation}
where we take $v$=1000 km s$^{-1}$ as the maximum velocity used to estimate the upper limits on the flux and $M_\mathrm{H_2,out}$ is the molecular hydrogen mass obtained from the flux (assuming $\alpha_\mathrm{CO} = 4$). Since we do not detect any emission from outflows, and the CO(2--1) emission is not resolved, we take the radius of the (undetected) outflow to be the size of the aperture used to extract the fluxes, $R_\mathrm{out}$ = 50 kpc ($\sim$ 6$"$ at $z=1-1.5$). The 3$\sigma$ upper limits on the mass outflow rates range from 130 to 660 $M_\odot$ yr$^{-1}$, with a median of 300 $M_\odot$ yr$^{-1}$, and are compiled in Table \ref{tab:sed}. The derived upper limits exceed the star formation rates derived for the sample. This is consistent with studies of outflows that find the mass outflow rate to be of the order of or below the star formation rate \citep{bischetti2019a,spilker2020}.

\subsection{Gas fractions and depletion times} \label{sec:tdep}

\begin{figure*}[!htb]
    \centering
    \includegraphics[scale=0.37]{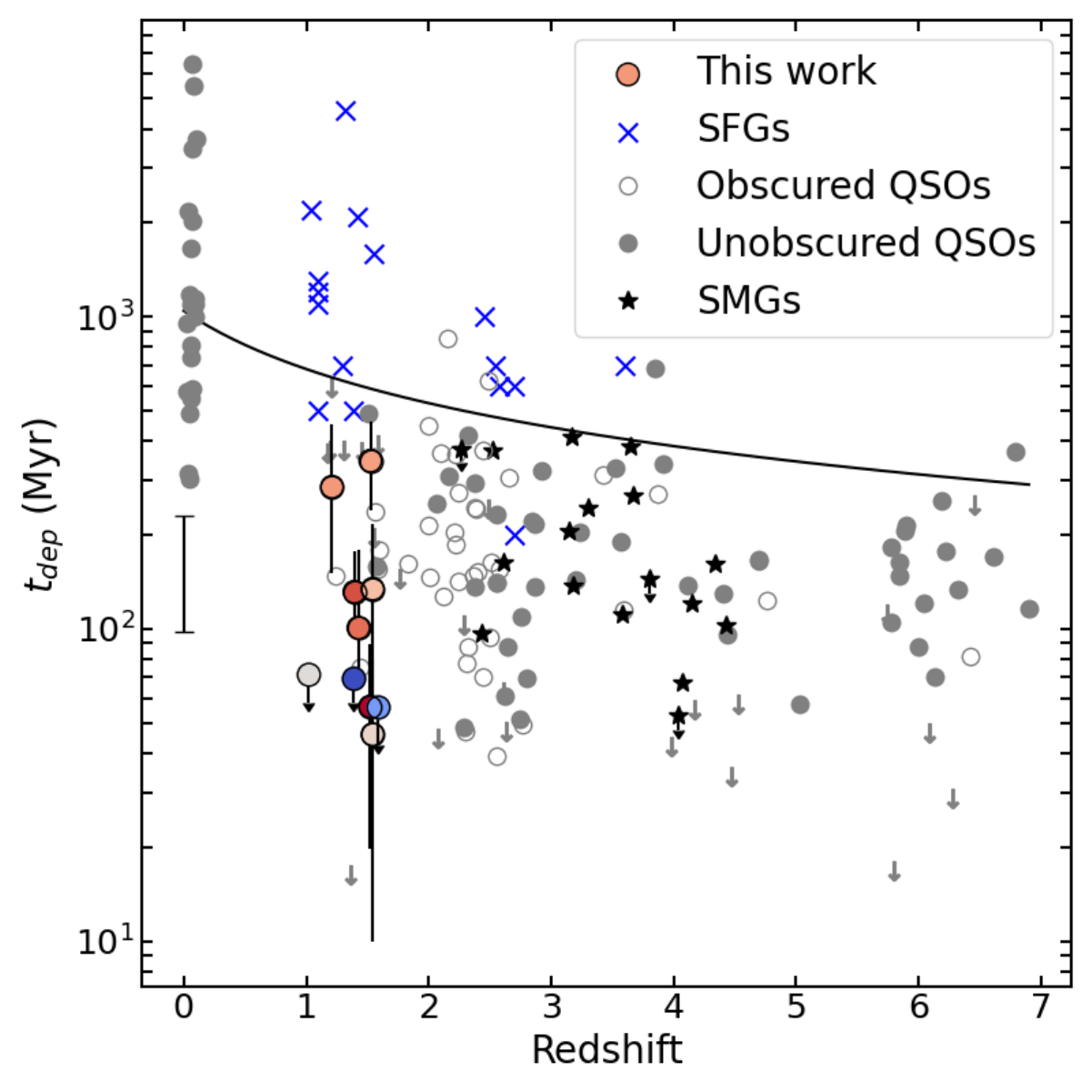}
    \includegraphics[scale=0.37]{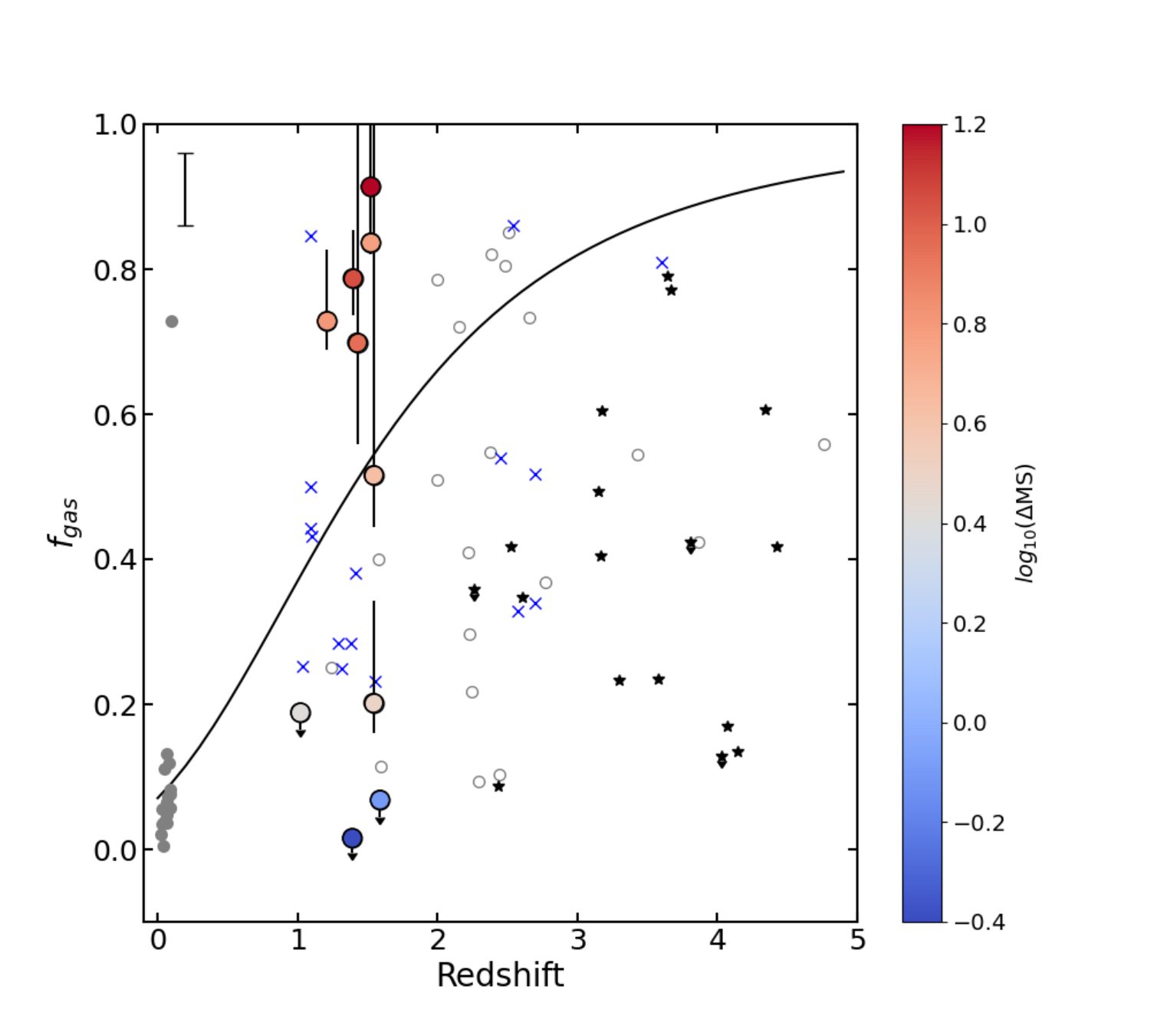}
    \caption{Left: Depletion time as a function of redshift for our sample, colour coded as a function of their offset from the MS. The comparison samples are as indicated in Fig. \ref{lco_v_z}, with their mean error shown by the vertical line at $z=0$. The solid line shows the expected trend \citep{tacconi2018} for a main--sequence galaxy of stellar mass 2$\times$10$^{9}$ M$_\odot$, the median of our sample. With depletion timescales in the range 45--900 Myr, most of our sample falls below the locus for normal, star--forming galaxies, and will quickly deplete the gas and transition into quiescence in the absence of gas replenishment. Right: Gas fraction ($M_\mathrm{gas}$/($M_\mathrm{gas}$+$M_\mathrm{*}$)) as a function of redshift. The high--redshift unobscured QSOs are not plotted as they lack estimates of their stellar masses. For the QSO comparison samples, the gas mass estimates have been adjusted for $\alpha_\mathrm{CO}$=4. On average, the detected QSOs are very gas--rich, while the non--detected QSOs, which are likely transitioning into quiescence, have almost depleted the available molecular gas. The largest gas fractions are found in the QSOs with the largest offset from the main--sequence, suggesting that the large gas mass is sustaining the high SFRs. }
    \label{fgas}
\end{figure*}

We can further explore two key parameters to understand the ISM conditions in our sample: the gas fraction and depletion time, defined as:
\begin{equation}
    f_\mathrm{gas} = \frac{M_\mathrm{gas}}{M_\mathrm{gas}+M_*}
\end{equation}
\begin{equation}
    t_\mathrm{dep} = \frac{M_\mathrm{gas}}{\mathrm{SFR}} =\frac{1}{\mathrm{SFE}} \ \mathrm{yr},
\end{equation}
where $M_*$ and SFR are as reported in Table \ref{tab:sed}. The gas fraction probes the amount of gas available for star formation, while the depletion time traces the time that it will take the galaxy to consume the available gas supply given its current star formation rate, assuming there is no gas replenishment.

 Fig. \ref{fgas} (left) shows the gas depletion times as a function of redshift colour coded as a function of offset from the main sequence, $\Delta$MS. We find depletion times in the range of 50--900 Myr. Generally, it appears the QSOs with larger offset from the MS have the shorter depletion times. The three non--detected QSOs show different behaviour, since they have the shortest depletion times and smallest MS offsets. This could be interpreted as further proof of their transitioning into quiescence -- they have low SFRs and short depletion times as they have exhausted their available molecular gas reservoirs. With a median $t_\mathrm{dep}$ = 90($\alpha_\mathrm{CO}$/4) Myr, our QSOs lie a factor of 7 below the locus of main--sequence galaxies \citep{tacconi2018}. This value is also lower than the higher--$z$ obscured and unobscured QSOs, which have a median $t_\mathrm{dep}$ of $\sim$170($\alpha_\mathrm{CO}$/4) Myr. Our sample also has lower $t_\mathrm{dep}$ than the higher--redshift SMGs, which have $t_\mathrm{dep}$ $\sim$150($\alpha_\mathrm{CO}$/1) Myr -- assuming $\alpha_\mathrm{CO}$ = 1 for the unobscured QSOs, as we do with the SMGs, would push their depletion times to 10--180 Myr, exacerbating the difference in timescales between both populations. The low--$z$ PG QSOs, however, have a much longer median $t_\mathrm{dep}$ of $\sim$1 Gyr, consistent with normal, star--forming galaxies. 

Fig. \ref{fgas} (right) shows the gas fraction as a function of redshift. The derived stellar masses (Table \ref{tab:sed}) suggest a wide range of $f_\mathrm{gas}$ from  $<$0.02 to 0.97 for our targets, assuming $\alpha_\mathrm{CO}$ = 4 ($<0.05$--0.90 if we assumed $\alpha_\mathrm{CO}$ = 1). The QSOs with the largest offset from the main sequence have the largest gas fractions, suggesting that a larger availability of molecular gas might be fueling the starbursts. The mean $f_\mathrm{gas}=0.67\pm0.22$ is comparable to what has been found in other high-$z$ QSOs \citep[e.g.][]{banerji2017,venemans2017,bischetti2021}, but larger than the gas fractions of SMGs (mean $f_\mathrm{gas}\sim0.4$, \cite{friascastillo2023}). There are still large uncertainties in these values due to the large errors of the stellar masses and the uncertain $\alpha_\mathrm{CO}$. Using the stellar masses derived from $M_\mathrm{BH}$ following \cite{ding2020} (see Section \ref{sec:sed}) yields equally large gas fractions (0.46--0.98), with a median $f_\mathrm{gas}$=0.77, for the QSOs where molecular gas is detected. Obscured QSOs also show a comparable spread in their gas fractions to those of our sample (\citealt{perna2018} found obscured QSOs to have lower f$_\mathrm{gas}$ than main--sequence galaxies, although this was driven by their assumption of $\alpha_\mathrm{CO}$=0.8). If we used $\alpha_\mathrm{CO}$=0.8 instead, our QSOs would have a mean $f_\mathrm{gas}$=0.4$\pm$0.2, in line with the average gas fraction found for SMGs at high--redshift. This highlights the large uncertainty regarding $\alpha_\mathrm{CO}$ for QSO--host galaxies, and the crucial need for high--resolution follow--up of the molecular gas in larger samples of QSOs to put stronger constraints on its value via, for example, dynamical modelling \citep{kakkad2017,Paraficz2018}.

\begin{figure*}[!htb]
    \centering
    \includegraphics[scale=0.37]{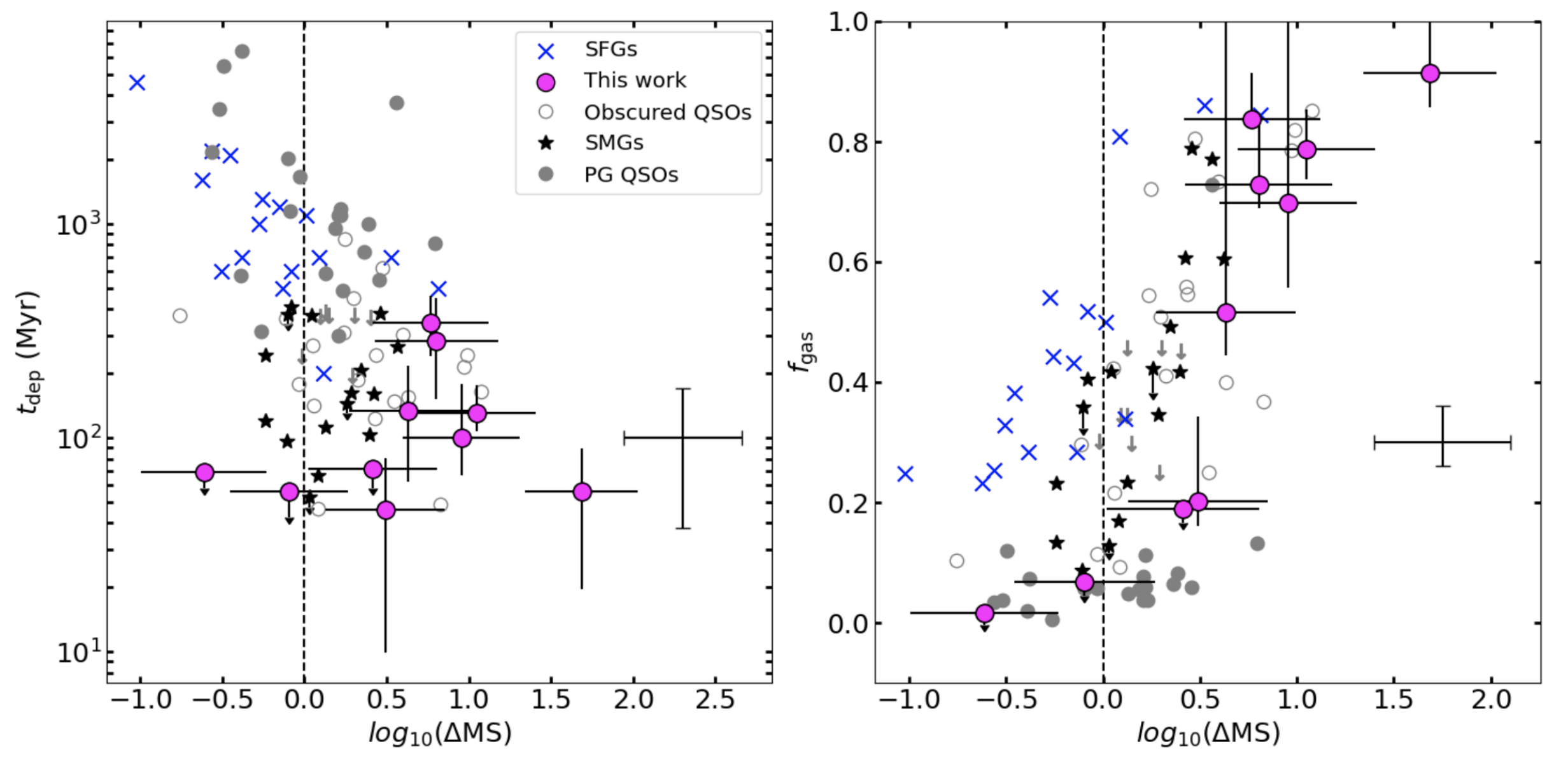}
    \caption{Depletion time (left) and gas fraction (right) as a function of offset from the main sequence ($\Delta$MS) as parameterised by \cite{speagle2014} for our QSOs (magenta circles). The comparison samples are as indicated in Fig. \ref{lco_v_z}, and we show their combined median error shown by a cross in each panel. For the QSO comparison samples, the gas mass estimates have been adjusted for $\alpha_\mathrm{CO}$=4. The vertical dashed line shows the position for sources on the main--sequence, $\Delta$ = 0. Our CO--detected QSOs follow the overall trend shown by the comparison samples of decreasing (increasing) depletion time (gas fraction) with increasing $\Delta$MS. B1608+656 and B1600+434 are outliers in this plot, as they have lower $t_\mathrm{dep}$ by about an order of magnitude than other star--forming galaxies and QSOs of similar $\Delta$MS, further supporting the idea that they are transitioning into quiescence.}
    \label{deltams}
\end{figure*}

In Fig. \ref{deltams} we show how the gas depletion times and gas fractions vary with offset from the main sequence for our sample of unobscured QSOs and the comparison samples from the literature. Our CO detected QSOs follow the decreasing trend seen in the $t_\mathrm{dep}$--$\Delta$MS plane. The non--detected QSOs, however, show depletion times about an order of magnitude lower than other SFGs and low--redshift QSOs at similar $\Delta$MS. This further indicates that these QSOs are transitioning into quiescence, and do not follow the relation derived for other star--forming galaxies. The gas fractions clearly increase with $\Delta$MS and follow within the error bars the trend shown by the literature samples. It is important to consider possible systematic uncertainties introduced by our choice of $\alpha_\mathrm{CO}$. Although we have tried to control for this by making consistent assumptions for all the QSO and SFG populations, we cannot rule out some level of systematic differences in $\alpha_\mathrm{CO}$ for our QSOs and those at high redshift, which may be significantly lower due to the starburst nature of some sources. The difference in the conversion factor could systematically shift our sample of QSOs to lower gas fractions and depletion times, increasing the difference with SFGs and SMGs.

\begin{figure*}[!hbt]
    \centering
    \includegraphics[scale=0.28]{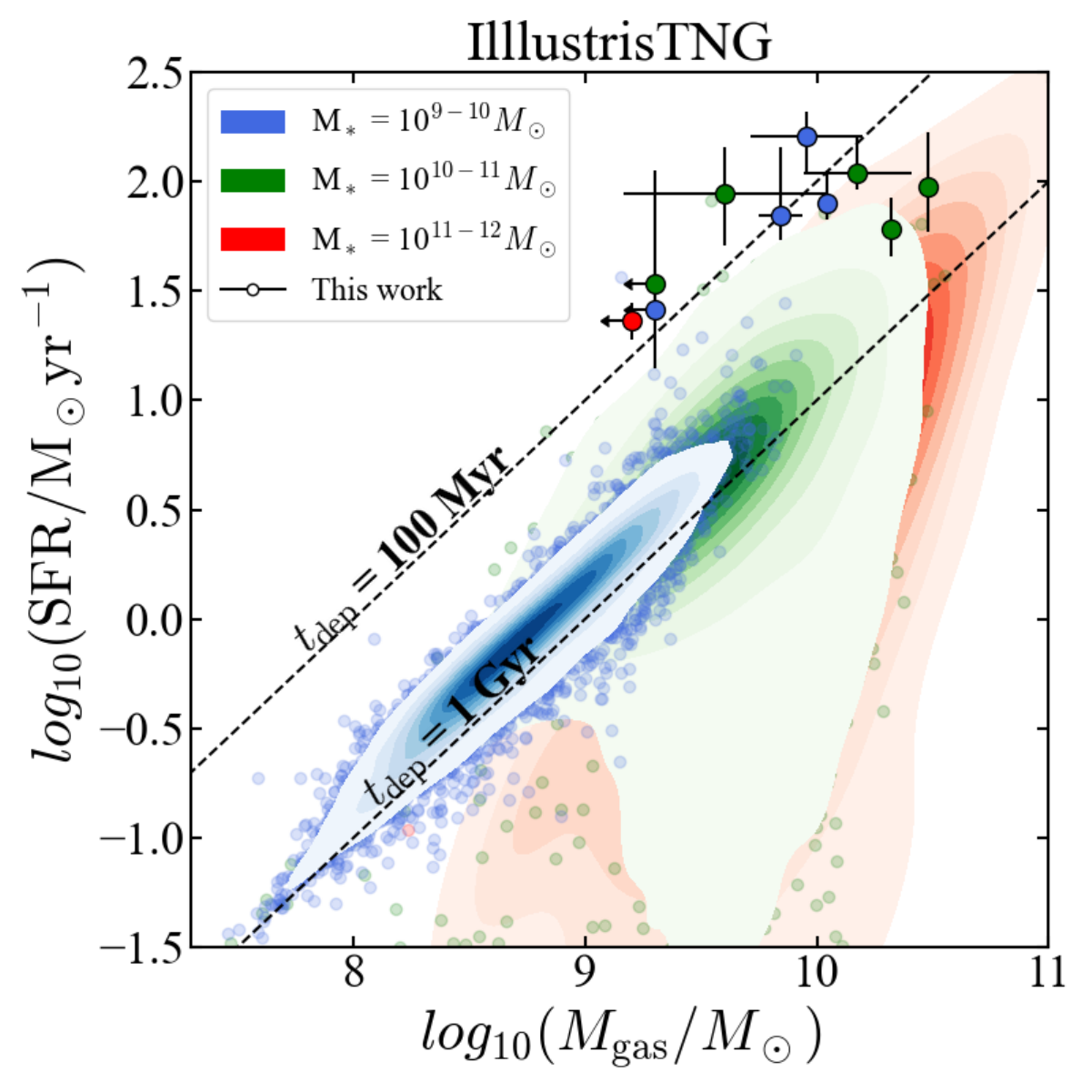}
     \includegraphics[scale=0.28]{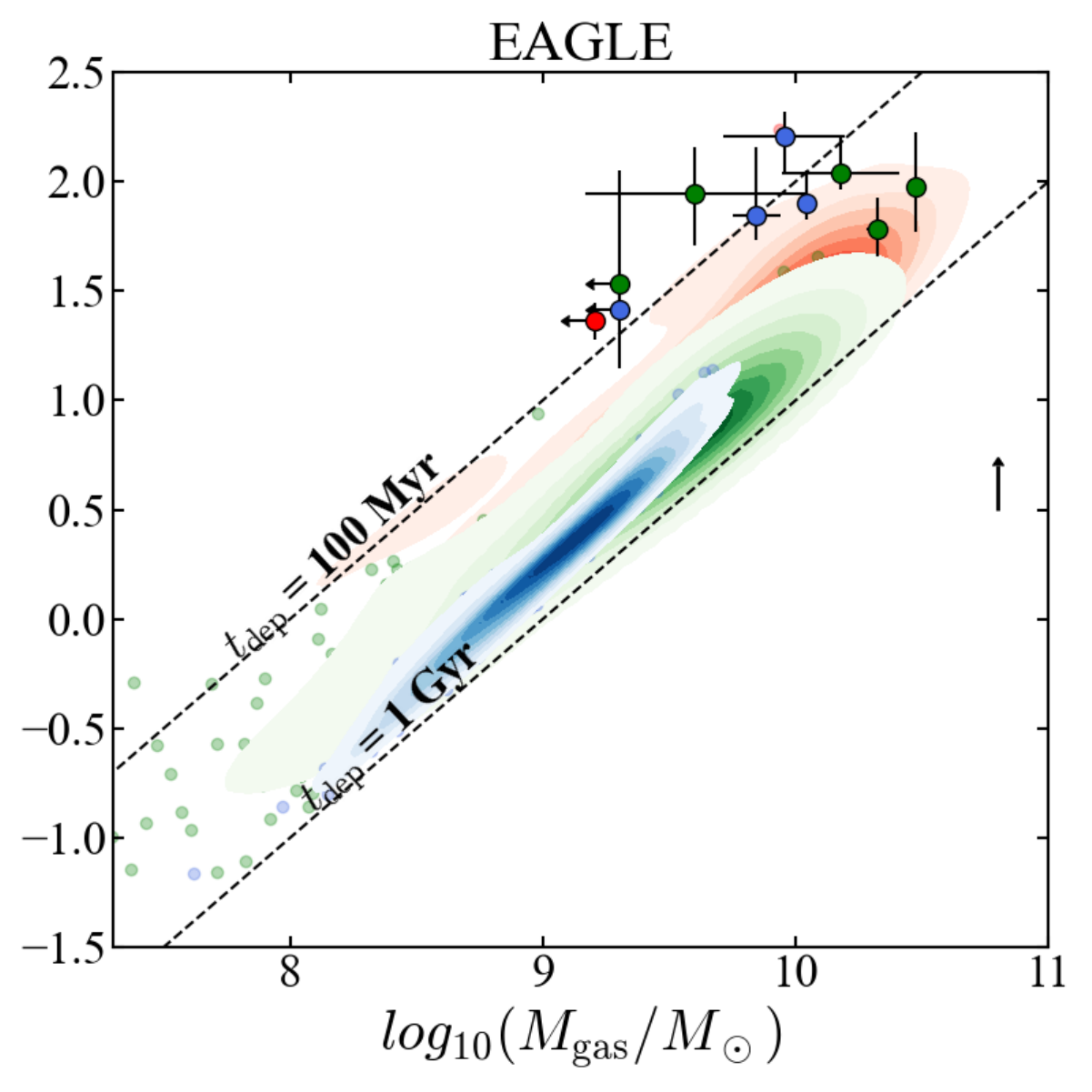}
      \includegraphics[scale=0.28]{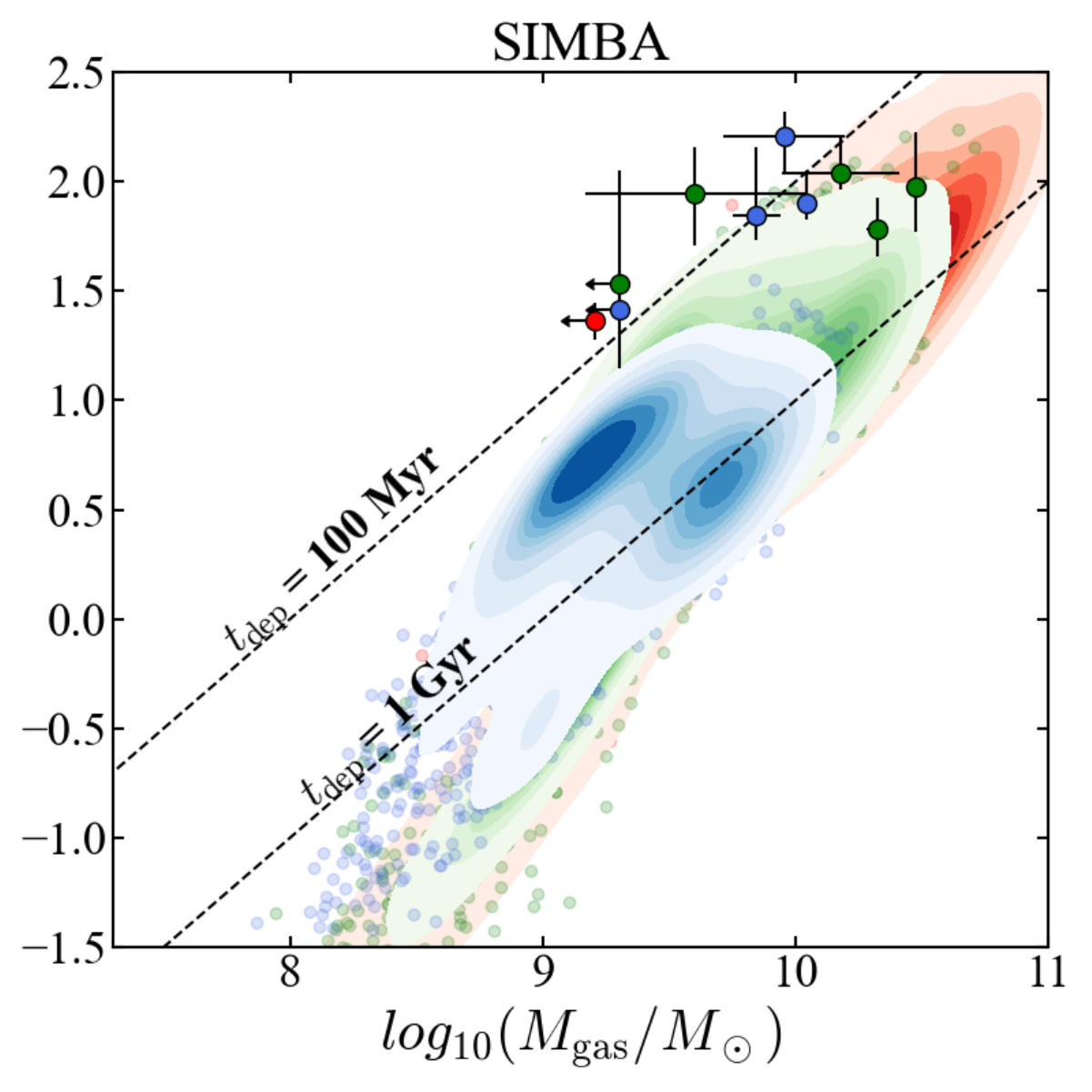}
    
    \caption{Distribution in the SFR--$M_\mathrm{gas}$ plane of the QSOs in this study, colour coded by their stellar mass. The contours show the position of galaxies with $L_\mathrm{bol} = $10$^{42-46}$ erg s$^{-1}$ and $M_{*}$ = 10$^{9-12}$ $M_\odot$ from the IllustrisTNG (left), EAGLE (middle) and SIMBA (right) hydrodynamical simulations, separated in bins of stellar mass (10$^{9-10} M_\odot$, blue; 10$^{10-11} M_\odot$, green; 10$^{11-12} M_\odot$, red). The dashed lines indicate depletion times of 100~Myr and 1~Gyr. Since the SFRs in EAGLE are known to be 0.2 dex lower than observations of main--sequence galaxies \citep{furlong2015,mcalpine2017}, we show by how much the SFRs would shift upwards if they matched the observations with the black arrow}. The simulations struggle to reproduce the high SFRs of our sample given their low stellar mass, which is likely due to a lack of sufficient resolution in the subgrid physics models. Despite using gravitational lensing to probe fainter systems, we are only probing the upper end of the predicted molecular gas mass distribution, suggesting that it will be challenging to detect non-lensed objects of similar gas masses.
    \label{fig:sims}
\end{figure*}

\subsection{Comparison with simulations} \label{sec:sims}

It is interesting to compare the results of our survey with predictions from the simulations. We select three of the current generation of hydrodynamic, cosmological simulations: EAGLE \citep{crain2015,schaye2015}, IllustrisTNG100 \citep{marinacci2018,naiman2018,nelson2018,pillepich2018,springel2018} and SIMBA \citep{dave2019}. All three simulations have boxes of similar size, L$\sim$100 cMpc (comoving Mpc). Following \cite{ward2022}, we select galaxies at $z=1$ with $L_\mathrm{bol}$ in the range 10$^{42-46}$ erg s$^{-1}$ and stellar masses in the range 10$^{9-12}$ $M_\odot$ in order to match the properties of our QSOs. We note that previous studies have shown that the SFRs in EAGLE are 0.2 dex lower than observations \citep{furlong2015,mcalpine2017}. Additionally, due to the limited volume of the simulations, EAGLE does not reach $L_\mathrm{bol}$ = 10$^{46}$ erg s$^{-1}$, while IllustrisTNG and SIMBA have very few galaxies in this range. This only affects comparison with J1633+3134, J1650+4251 and B1152+434, which have the highest $L_\mathrm{bol}$.

In Fig. \ref{fig:sims} we show the SFR--$M_\mathrm{gas}$ plane, where we compare our QSOs with galaxies from the three simulations at $z=1$, separated in bins of stellar mass. Despite the boost in signal provided by gravitational lensing, we are only able to probe the upper end of the molecular gas mass distribution with our QSOs. This suggests that it will be challenging to detect non--lensed systems of comparable mass with our current facilities.

Looking at the locus of our QSOs in the SFR--$M_\mathrm{gas}$ plane, we find that most of their SFRs are underestimated in every simulation, leading to an overestimation of their depletion timescales. This mismatch in SFR values would only worsen if we assumed a lower $\alpha_\mathrm{CO}$, which would decrease both the molecular gas mass and the depletion times. Although the simulations do have galaxies with SFRs comparable to those of the QSOs, those values are only expected for galaxies of stellar mass $M_*\sim$10$^{11} M_\odot$. One possible explanation is that \textsc{AGNfitter} did not properly remove all of the AGN contribution to $L_\mathrm{IR}$. However, we do not expect this to decrease the IR luminosities by more than 0.1 dex \citep{Kirkpatrick2019}, which would still not be enough to solve the discrepancy. This tension in SFRs is more likely the result of the difficulty that simulations still have in producing galaxies with elevated SFRs. The lack of starburst galaxies in current hydrodynamical simulations (EAGLE, SIMBA, Illustris, Horizon-AGN) has been pointed out in previous studies of galaxy formation models \citep[e.g.][]{sparre2015,katisanis2017,rinaldi2022}. This may be due to the insufficient resolution of galaxy models, which cannot properly model the multiphase ISM, in particular the cold phase where star formation takes place. Star formation in simulations follows the Kennicutt-Schmidt law \citep{kennicutt1998}. However, in order to produce a starburst, very high densities only reached in the cold phase of the ISM are necessary. But if the molecular gas gets too cold and dense, the Jeans mass and length become smaller than the resolved mass of the simulation. Therefore, to prevent this from happening, a pressure floor is imposed on the simulation, thereby preventing the gas from becoming cold and dense enough to produce a starburst.

\section{Discussion}
\label{sec:discussion}

In the canonical picture of massive galaxy evolution \citep{sanders_mirabel1985,alexander2005,hopkins2008}, unobscured QSO hosts are the end product of the merger between two gas--rich galaxies, and are expected to be systematically gas--poorer than `normal' galaxies with the same stellar mass and have little remaining star formation due to QSO feedback. Many studies have explored the gas content in QSO hosts compared to non--QSO systems, both at low \citep[e.g.][]{rosario2018,jarvis2020,shangguan2020,ramosalmeida2022} and high redshift \citep[e.g.][]{kakkad2017,perna2018,damato2020,Kirkpatrick2019,bischetti2021}. While these studies are very heterogeneous, there is a growing consensus that QSO host galaxies are as gas--rich as their non--QSO counterparts, a conclusion supported by state--of--the-art simulations \citep{ward2022}. This however depends on the highly uncertain $\alpha_\mathrm{CO}$ value for QSOs, to which the gas fraction is very sensitive.

We have detected CO(2--1) line emission in 70\% of our sample of FIR--bright, unobscured QSOs (Figs. \ref{detections}, \ref{detections_cont}). The derived gas masses of $<$1.6--36 ($\alpha_\mathrm{CO}$/4) $\times$ 10$^{9}$ M$_\odot$ are about an order of magnitude below those of normal, star--forming galaxies (Fig. \ref{lco_v_z}). Our SED analysis reveals that these systems also have lower stellar masses than the average population of obscured and unobscured high--redshift QSOs, which we are able to detect thanks to gravitational lensing. These are therefore likely to be a younger and/or less massive population than the high--redshift QSOs studied so far. Nevertheless, we infer gas fractions between $<$2\% and 97\% (for $\alpha_\mathrm{CO}$=4, Fig. \ref{fgas}), although with large errors introduced by uncertainties in the stellar mass. Taken at face value, our sample thus spans the entire physical range of $f_\mathrm{gas}$, including both very gas--rich and gas--poor galaxies. However, five of our targets are conspicuously gas--rich, with $f_\mathrm{gas} >$ 50\%. These gas fractions are comparable to --or exceeding-- typical main sequence galaxies at their redshift and stellar mass (for example, ASPECS MS galaxies in Figs. \ref{main_sequence} and \ref{fgas}), and support the results from previous studies pointing towards QSO--host being as gas--rich as non--QSO hosts \citep[e.g.][]{kakkad2017,rodighiero2019,damato2020,jarvis2020,shangguan2020}. Recent cosmological simulations also predict that luminous AGN are associated with gas--rich star--forming galaxies \citep{ward2022}. This co--existence of star formation and AGN activity \citep{floyd2013,stanley2017,symeonidis2022} is likely caused by the supply of cold molecular gas driving both phenomena. Nevertheless, high gas fractions do not mean that AGN feedback is not affecting the gas reservoirs of the galaxies, but, considering the short timescales of QSO activity, it becomes challenging to detect their impact on $f_\mathrm{fgas}$. Instead, it is likely that the cumulative effects of several episodes of AGN activity are necessary to significantly depress the gas fractions \citep{piotrowska2022}, and could indicate that the central AGN in B1608+656, B1152+200 and B1600+434 have been active for a longer period of time compared to the rest of the sample.

Despite the large gas fractions, the high star formation rates of our targets imply that the available molecular gas will quickly be depleted, and star formation will subsequently quench. The depletion times of $\sim$90 Myr (or $\sim$25 Myr for $\alpha_\mathrm{CO} = 1$, Fig. \ref{fgas}) suggest that these galaxies will quench and turn into gas--poor unobscured QSOs during the lifetime of the quasar, rather than much shorter timescales of $\sim$1 Myr, as had been suggested by \cite{simpson2012}. Such short depletion times have been routinely found in other high--redshift QSOs \citep[e.g.][although see \citealt{Paraficz2018}]{riechers2011,bischetti2021}, and are significantly below those of low--redshift QSOs, with $t_\mathrm{dep}\sim$1 Gyr.

In summary,  the current data for our sample of QSOs suggests that the high star formation rates are enough to explain the deficit in CO luminosity (Fig. \ref{sfe}) and short depletion times (Figs. \ref{fgas}, \ref{deltams}). While this does not rule out the presence of feedback, current or past, from the central AGN which might be falling bellow the detection limit of our NOEMA observations, the high gas fractions and lack of evidence for cold gas outflows suggest however that, if present, AGN feedback has not significantly impacted the gas reservoirs yet. Nevertheless, outflows observed in luminous QSOs typically correspond to $\sim$2\% of the peak of the CO emission line profiles \citep[e.g.][]{feruglio2017}, so deeper CO(2--1) observations would be required to determine the prevalence of ongoing outflows in these targets. Along with higher angular resolution observations of low $J$-level CO and FIR emission to probe the size and distribution of the cold gas and dust reservoirs, these diagnostics will be useful to determine the relative importance of AGN and stellar feedback in the process of quenching in QSO--host galaxies. Higher angular resolution observations would also help study the impact of galaxy morphology the molecular gas content of our sample of QSOs, as \cite{husemann2017} and \cite{ramosalmeida2022} found that disc--dominated and merging QSOs have larger gas masses than bulge--dominated QSOs.

Finally, B1608+656, B1152+200 and B1600+434 have not been detected in CO emission. These are three of the four radio--loud QSOs in our sample, and show bright continuum ($\lambda_\mathrm{rest}$ = 1.3mm) emission likely arising from the central AGN and have molecular gas masses $M_\mathrm{gas}<10^9$ $M_\odot$. These radio jets are likely to cause jet--mode AGN feedback that will impact the gas--reservoirs of these radio--loud, QSO--host galaxies, possibly over a long period of time \citep[e.g.][]{mukherjee2018,fotopoulou2019,jarvis2019,zovaro2019,couto2023,tamhane2023}. MG 0414+0534 is another radio--loud, unobscured QSO detected in mid--$J$ CO emission \citep{barvainis1998} but not in low--$J$ emission \citep{sharon2016}. Conversely, radio--loud but obscured QSOs such as B1938+666 \citep{sharon2016} or MG 0751+2716 \citep{riechers2011} do show low--excitation gas reservoirs detected in CO(1--0). If obscured and unobscured radio--loud QSOs are evolutionary connected, this would be consistent with the radio--loud, obscured phase occurring earlier, until eventually the heat and turbulence injected by the radio jets reduce the cold molecular gas and star formation halts. A larger, well--defined sample of both obscured and unobscured, radio--loud QSOs observed in low--$J$ emission will be needed to explore this scenario.

\medskip

\section{Conclusions}
\label{sec:conclusions}

We have presented NOEMA observations of CO(2--1) line and continuum emission in a sample of ten gravitationally lensed, unobscured QSOs at $z =$ 1--1.5. These observations significantly increase the number of unobscured QSOs detected in CO line emission in this redshift range. Our main conclusions are as follows:

- We detect CO(2--1) in seven of our targets (70\% detection rate), with line luminosities of 0.9--7.7 $\times$ 10$^{9}$ $\mathrm{K \ km \  s^{-1} \ pc^2}$. For the three QSOs not detected, we place 3$\sigma$ upper limits on their line luminosities of $\leq$0.4 $\times$ 10$^{9}$ $\mathrm{K \ km \  s^{-1} \ pc^2}$. We also detect far--infrared continuum emission underlying the CO line emission in four of the CO--detected targets. Three out of the four radio--loud QSOs in our sample show no CO emission but have very strong (S/N$>$200) continuum emission, likely originating from the radio jets. Although we currently do not find evidence of outflows in any of the targets detected in CO(2--1), more data is needed to definitively rule out significant on-going outflow activity contributing to further gas depletion in the QSOs.

- We collect the available photometry for the ten targets and separate the emission coming from the lens and lensed galaxies. We then fit the SED of the lensed QSOs and their host galaxies using \textsc{AGNFitter}. We find a large range of stellar masses and SFRs which, according to empirical scaling relations, classify six of the targets as starbursts ($\Delta$MS$>$0.6 dex), with three more consistent with being on the main sequence of star--forming galaxies. Our study supports the idea that AGN and starburst activity co--exist in the host galaxies of FIR--bright, unobscured QSO. If there is any significant SFR suppression by AGN feedback, it must occur on longer timescales. The three radio--loud QSOs not detected in CO, however, fall on or below the main--sequence, which suggests that they are transitioning into quiescence.

- Only three of our targets fall within the scatter of the $L'_\mathrm{CO}$--$L'_\mathrm{IR}$ empirical relation for either star--forming or starburst galaxies. The seven remaining QSOs fall well below this relation. This could be a sign of AGN or star formation feedback either depleting the molecular gas reservoir or heating up the gas and thus preventing it from collapsing to form new stars. We find the largest outliers from the $L'_\mathrm{CO}$--$L'_\mathrm{IR}$ relation to be the three non--detected QSOs. Along with their position on the main--sequence, short gas depletion times and low gas fractions, we interpret this as a possible sign of jet--mode AGN feedback contributing to the depletion of the gas reservoirs in these sources.

- We find total cold molecular gas masses in the range $\leq$2--30 $\times$ 10$^{9}$ ($\alpha_\mathrm{CO}$/4) $M_\odot$. This implies high gas fractions of $\sim$50\% (for $\alpha_\mathrm{CO}$=4) for the detected galaxies. Despite these targets being gas--rich, the high SFRs result in a range of depletion times of 50--900 Myr, with a median of only $\sim$90 Myr, a factor of 7 below the expected value for main--sequence galaxies. These short depletion times are only reinforced if we assume a starburst--like conversion factor of $\alpha_\mathrm{CO}$ = 1. Our QSOs are thus undergoing a period of intense growth, and will quickly turn the available gas into stars, subsequently quenching star formation.

- We compare our QSOs with galaxies from the hydrodynamical simulations IllustrisTNG100, EAGLE, and SIMBA, in the same redshift, stellar mass and bolometric luminosity intervals. We show that, while our sample is comparable in total gas mass to the most gas--rich galaxies selected, the simulations struggle to reproduce their high SFRs and thus overestimate their gas depletion times, especially when we compare in smaller bins of stellar mass. This lack of starbursts is linked to the fact that the current simulations do not have the necessary resolution to successfully model the small scale physics involved in the starburst phenomenon, as has been previously noted for the bulk of the simulated galaxy population.

The NOEMA observations analysed in this work provide important information about the properties of the cold ISM of FIR--bright, unobscured QSOs just after the peak epoch of QSO activity and massive galaxy assembly. We find that these systems lie preferentially in low--mass, gas--rich host galaxies undergoing starbursts. This phase of intense growth will use up the molecular gas reservoir in a few hundreds of Myr. Follow-up observations with higher angular resolution are needed to map the cold gas kinematics in their hosts and establish the presence of merger signatures to better understand the cause of the intense starburst, as well as to establish the importance of AGN feedback in the depletion of the gas reservoirs.

\begin{acknowledgements}
We would like to thank the anonymous referee for their constructive comments that helped significantly improve this work. This work is based on observations carried out under projects S19CC and W20CM with the IRAM NOEMA Interferometer. IRAM is supported by INSU/CNRS (France), MPG (Germany) and IGN (Spain).
M. F. C. and M. R. thank IRAM and J. M. Winters in particular for hosting the online data reduction.\\
M. R. is supported by the NWO Veni project `Under the lens' (VI.Veni.202.225).
M. F. C., M. R. and J. H. acknowledge support of the VIDI research programme with project number 639.042.611, which is (partly) financed by the Netherlands Organisation for Scientific Research (NWO). J.H. acknowledges support from the ERC Consolidator Grant 101088676 (VOYAJ). CMH acknowledges funding from an United Kingdom Research and Innovation grant (code: MR/V022830/1. SRW acknowledges funding from the Deutsche Forschungsgemeinschaft (DFG, German Research Foundation) under Germany’s Excellence Strategy: EXC2094-390783311. This work is based on the research supported in part by the National Research Foundation of South Africa (Grant Numbers: 128943). The research leading to these results has received funding from the European Union's Horizon 2020 research and innovation program under grant agreement No 730562 [RadioNet]. The NASA/IPAC Extragalactic Database (NED) is funded by the National Aeronautics and Space Administration and operated by the California Institute of Technology.

\bigskip

\end{acknowledgements}

\bibliographystyle{aa} 
\bibliography{aanda} 

\begin{appendix}

\section{Observations summary} \label{appendix:obs_summary}

\begin{table*}[!htbp]
\caption{Summary of NOEMA observations: target ID, date of observations, antenna configuration, total and on-source time ($t_\mathrm{tot}$ and $t_\mathrm{on}$). \label{tab:obs} }
\begin{center}
 \begin{tabular}{@{}lcccccl @{}}
 \hline \hline
Target & Date & Configuration & $t_\mathrm{tot}$ & $t_\mathrm{on}$ & pwv & Notes \\
 & & & [h] & [h] & [mm] & \\
\hline
J1524+4409 &  2019 June 10 & 9D & 2.6 & &\\
B1608+656 &  2019 June 26 & 9D & 2.2 & & 6 -- 8 & \\ 
 &  2019 June 28  & 9D & 2.2 & & 3 -- 6 & Data discarded.\\ 
 &  2019 June 29 & 9D & 5.2 & & 5 -- 6 & \\ 
J1330+1810 & 2019 Sept 1 & 9D & 1.9 & & & Flux calibration bootstrapped.\\
& 2019 Dec 28 & 10C & 1.5 & & & New baseline solution applied; \\
& & & & & & A08 shadowed towards the end of the track.\\
J1455+1447 & 2019 Aug 29 & 9D & 3.4 & & \\
& 2019 Oct 06 & 9D & 2.6 & \\
& 2019 Oct 07 & 9-special & 1.5 & & & \\
& 2019 Oct 12 & 9-special & 3.4 &\\
J1633+3134 & 2019 July 3 & 9D & 2.2 & & \\
& 2019 Aug 26 & 9D & 0.8 & \\
& 2019 Aug 27 & 9D & 1.9 & & & Data discarded.\\
& 2019 Aug 28 & 9D & 2.2 &\\
& 2019 Sept 3 & 9D & 1.1 & \\
J0924+0219 & 2019 Jun 8 & 9D & 3 & & & \\
&  2019 Sep 25 & 9D & 3 & & & \\
&  2019 Sep 30 & 9D & 0.8 & & & \\
&  2020 Jan 21 & 10C & 1.1 & & \\
J1650+4251 & 2019 Aug 4 & 9D & 3.0 & & 4 -- 8\\
 & 2019 Aug 5 & 9D & 1.5 & & 9 -- 8 \\
B1152+200 & 2020 Dec 30 & 11C & 2.2 & &\\
B1600+434 &  2020 Nov 28 & 10-special & 1.1 & &\\
& 2020 Dec 05 & 9-special& 1.5 & &\\
& 2020 Dec 22 & 11C & 1.9 & &\\
J0806+2006 & 2020 Dec 23 & 11C & 1.9 & &\\
& 2020 Dec 30 & 11C & 1.5 & & \\
\hline
 \hline
 \end{tabular}
\end{center}
\end{table*}

\section{SED fits} \label{appendix:sed_fits_app}

\begin{figure*}[!htbp]
    \centering
    \includegraphics[scale=0.4]{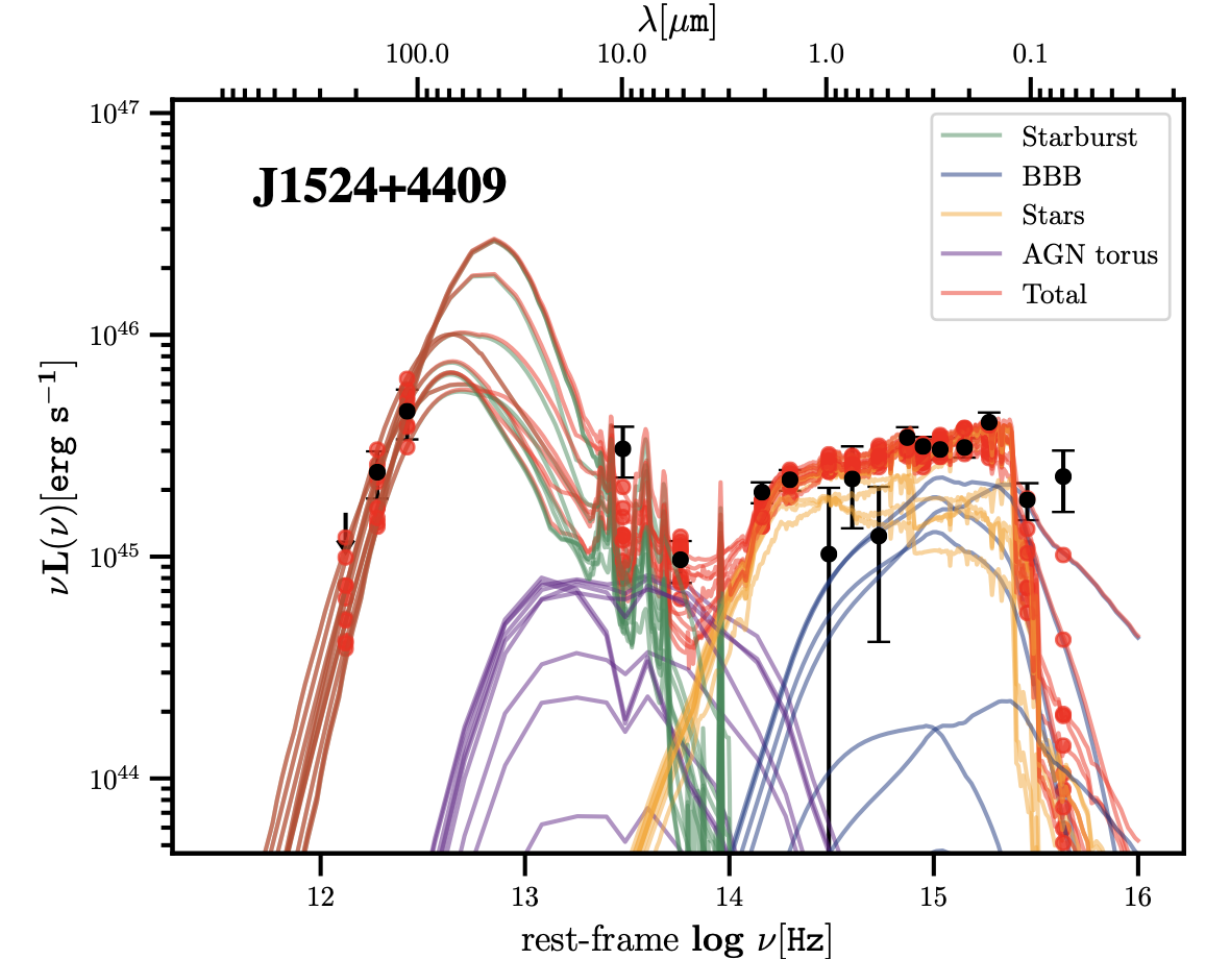}
    \includegraphics[scale=0.4]{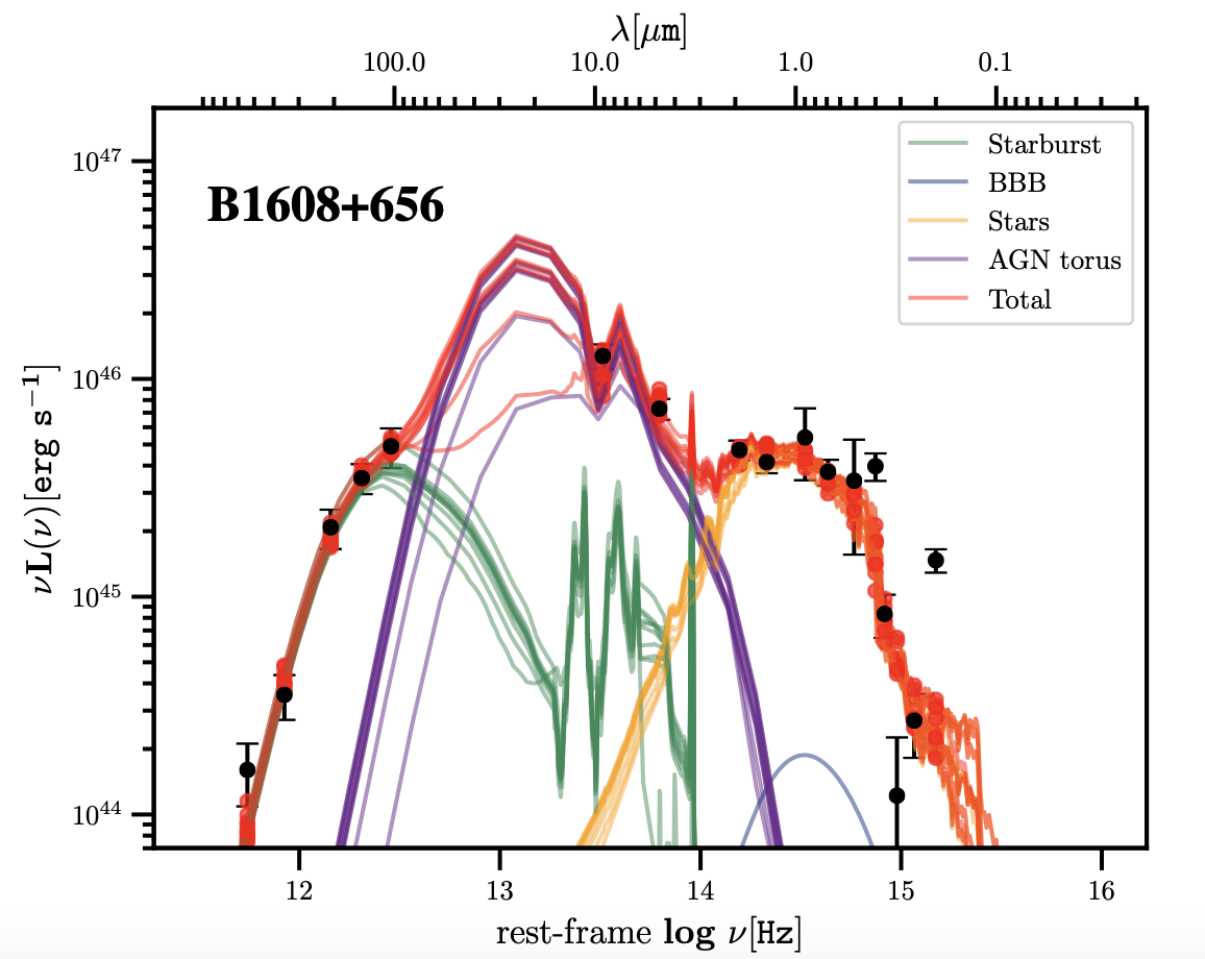}
    \includegraphics[scale=0.4]{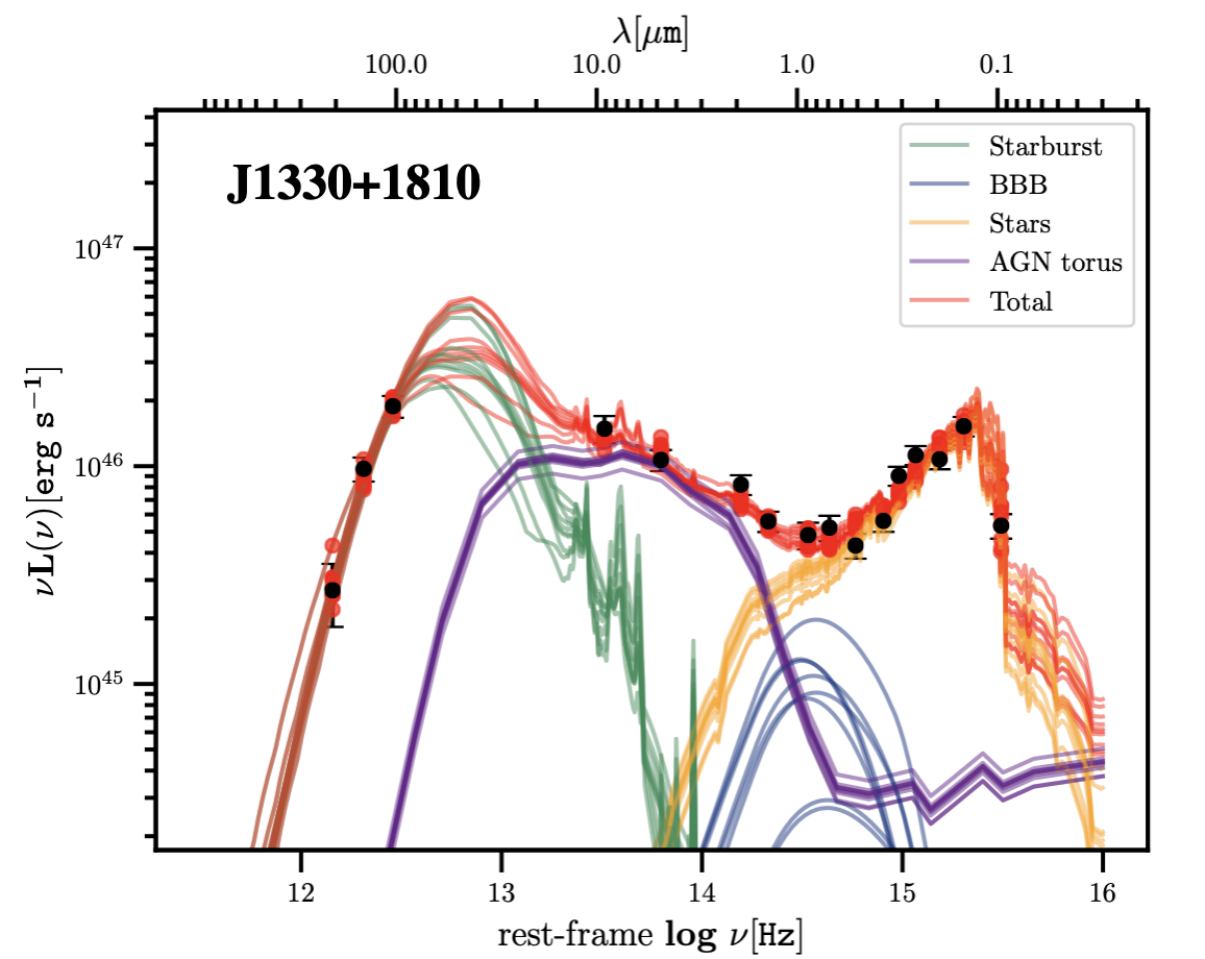}
    \includegraphics[scale=0.4]{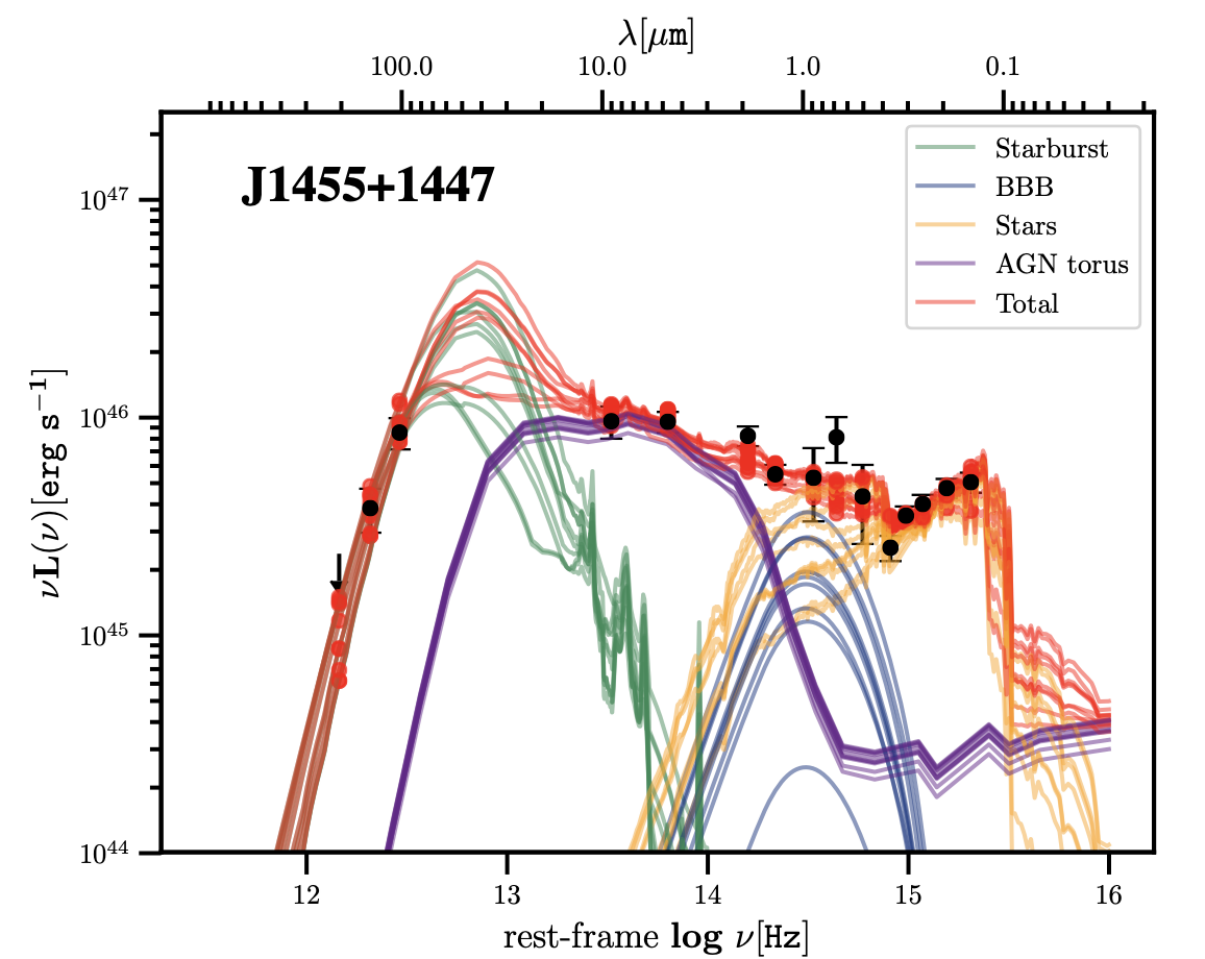}
    \includegraphics[scale=0.4]{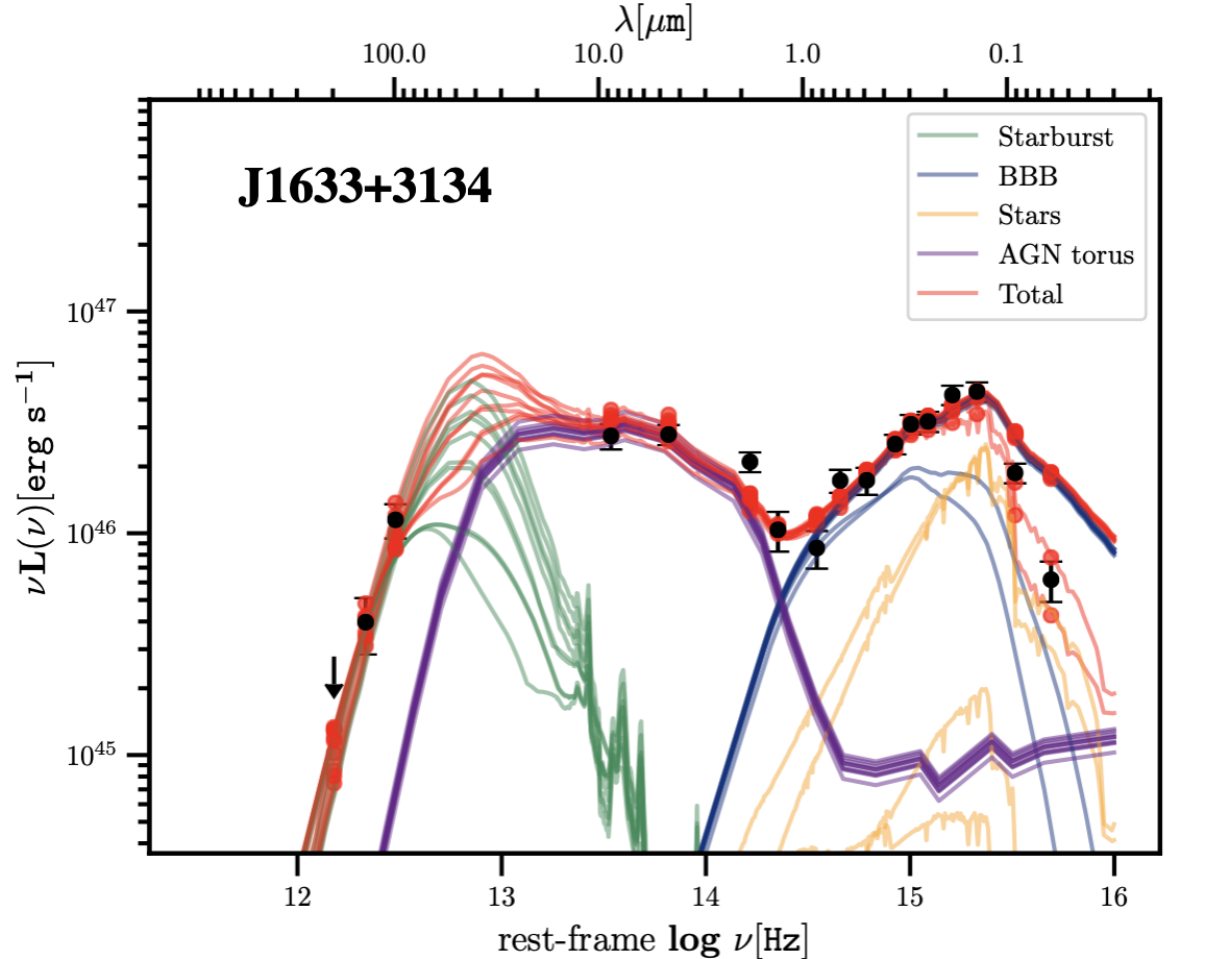}
    \includegraphics[scale=0.4]{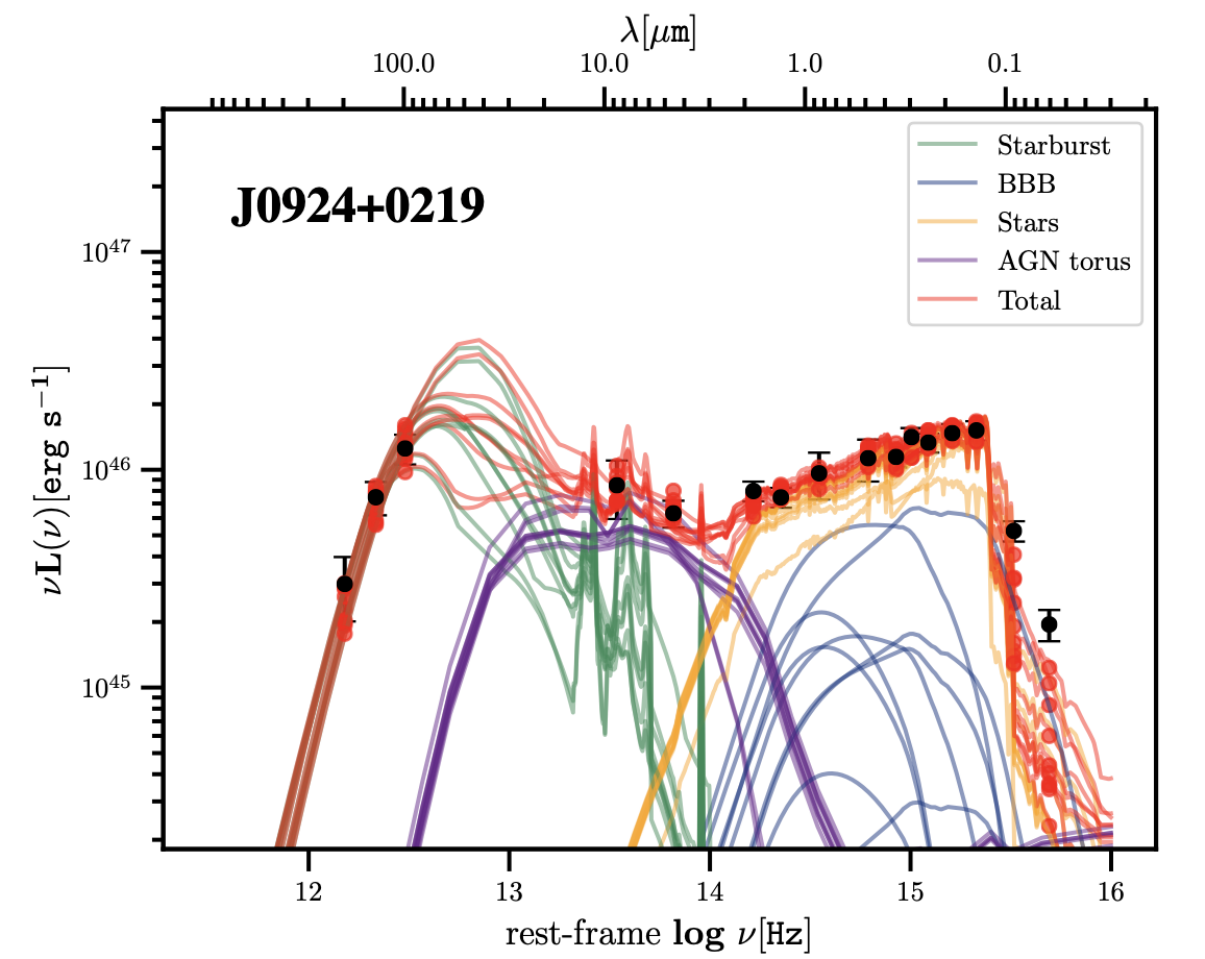}
    \caption{From top left to bottom right: SED fitting of J1524+4409, B1608+656, J1330+1810, J1455+1447, J1633+3134, and J0924+0214. The black dots represent the observed data points; 3$\sigma$ upper limits are shown as downward arrows. The best-fit SED obtained with AGNfitter \citep{Calistro2016}, as well as the different components used to model the total flux are shown as labelled in the figure. Ten realisations picked from the posterior distribution are shown.}
    \label{fig:sed1}
\end{figure*}

\begin{figure*}[!htbp]
    \centering
    \includegraphics[scale=0.4]{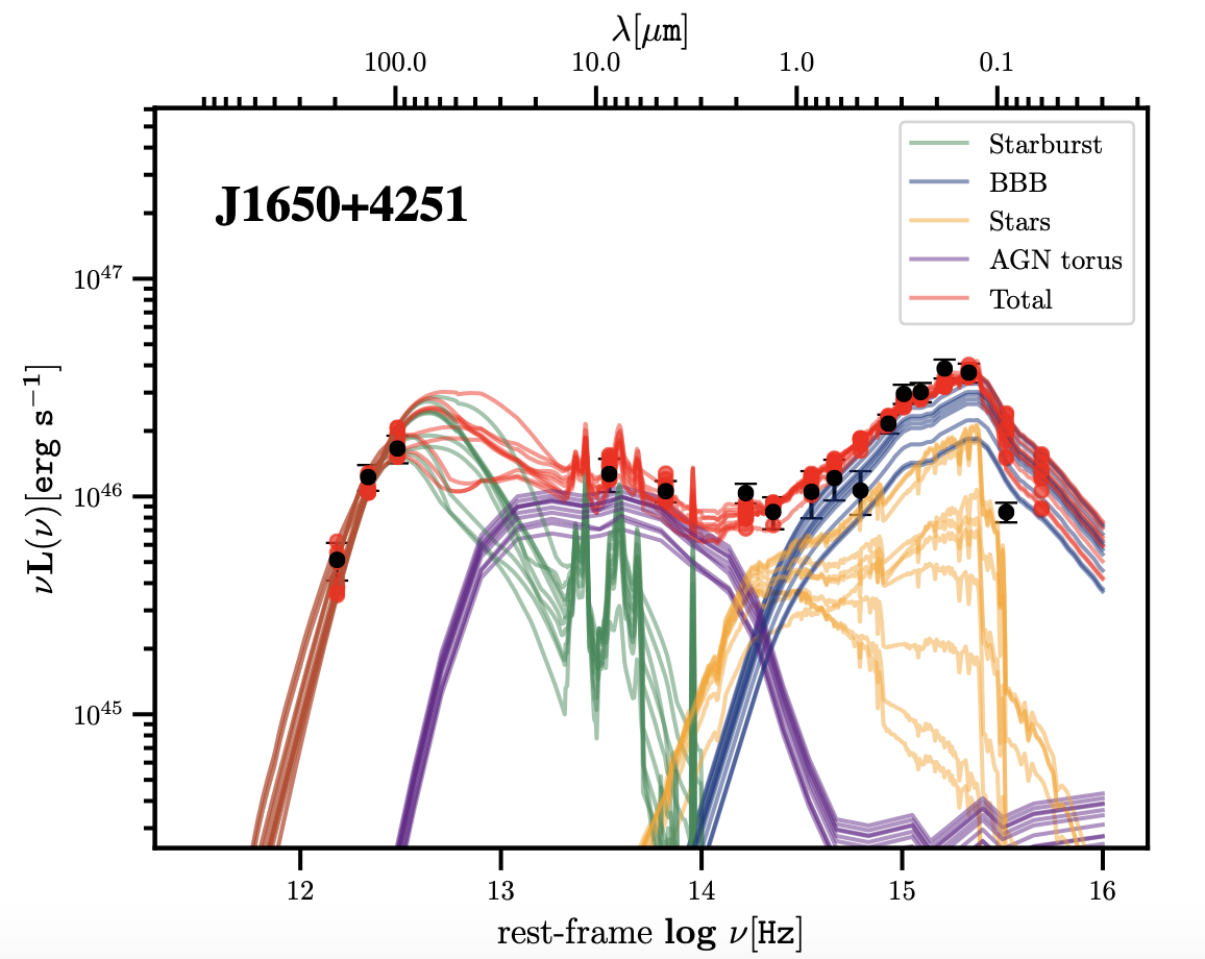}
    \includegraphics[scale=0.4]{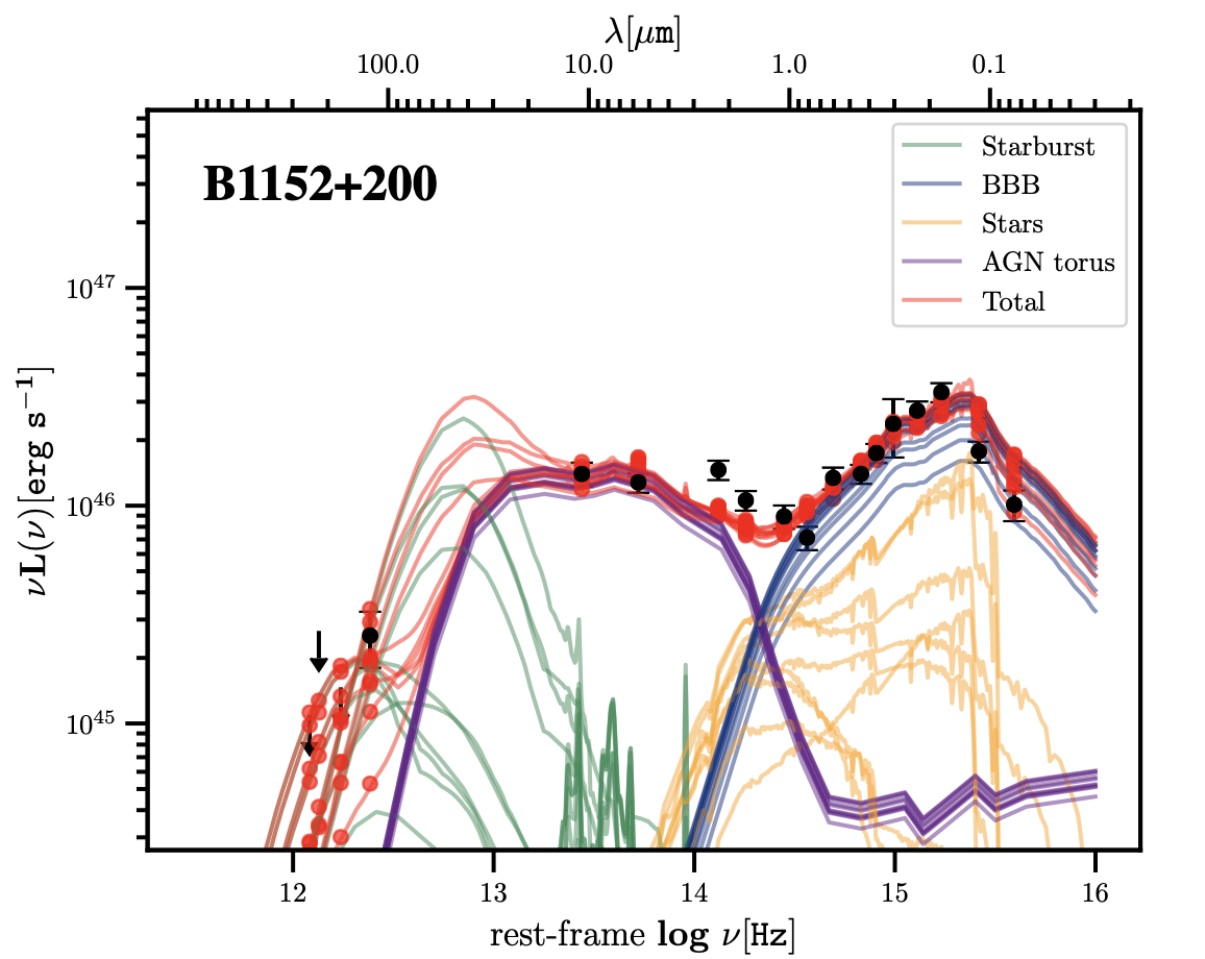}
    \includegraphics[scale=0.4]{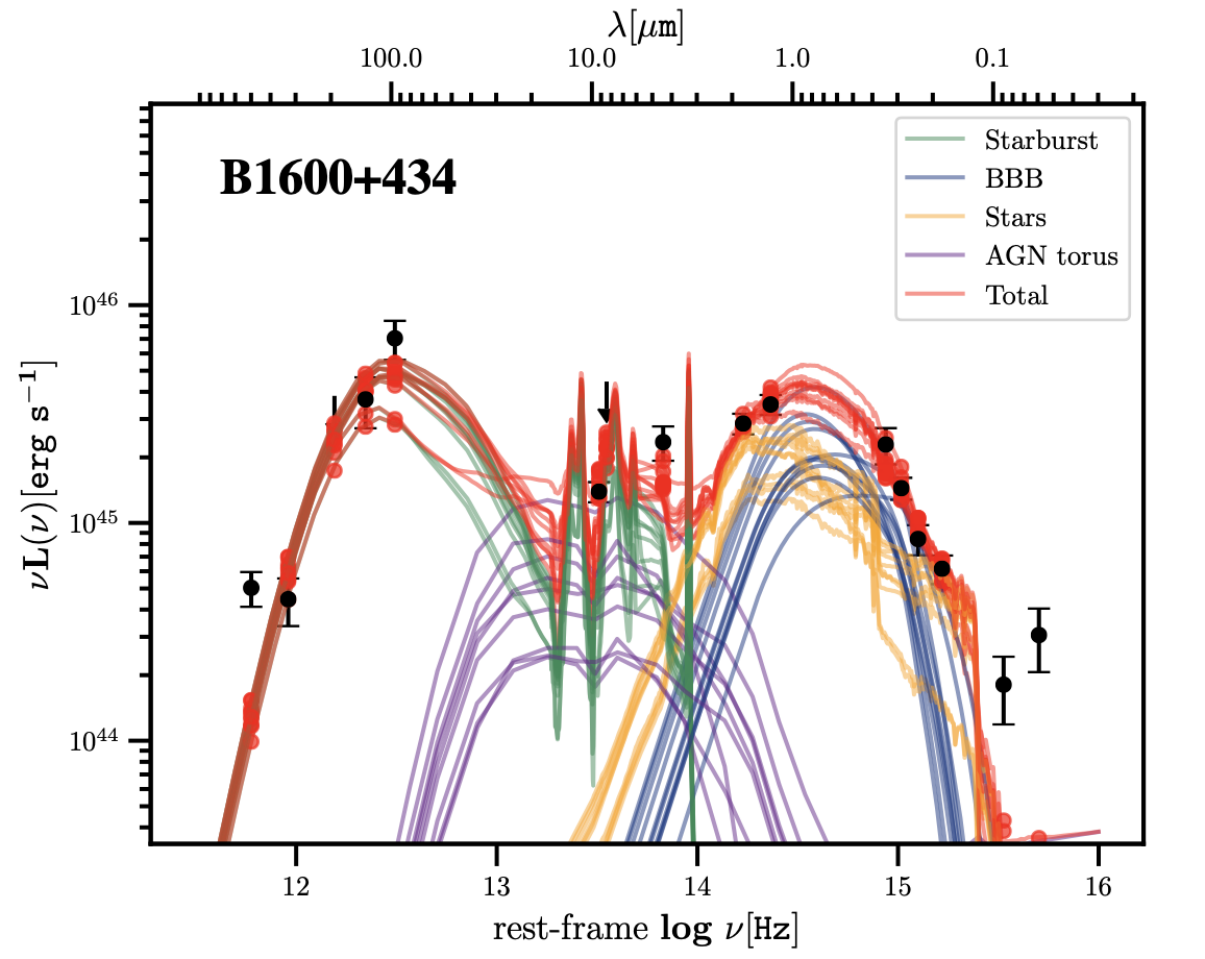}
    \includegraphics[scale=0.4]{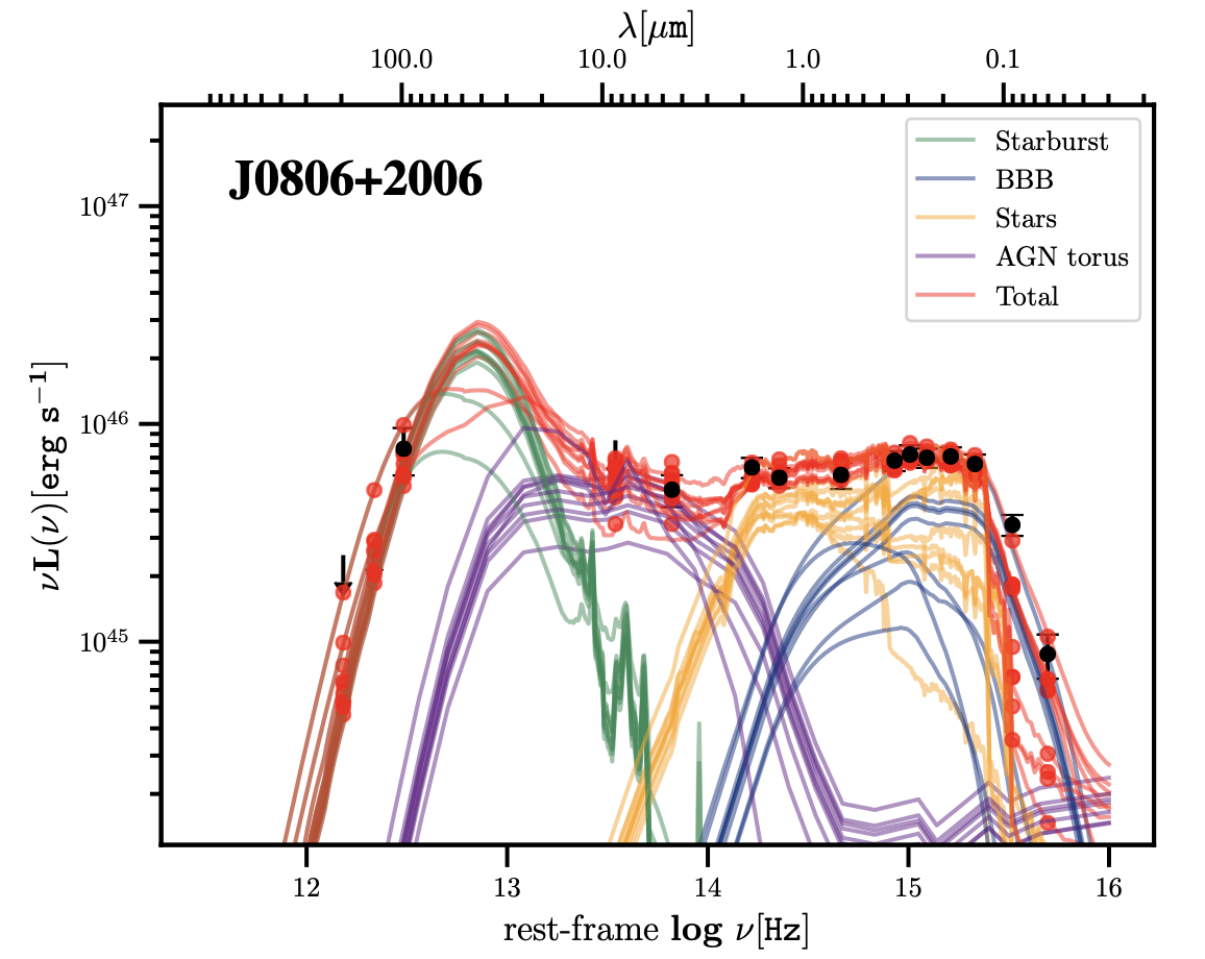}
    
    \caption{Fig. \ref{fig:sed1} continued. From top left to bottom right: SED fitting of J1650+4251, B1152+200, B1600+434, and J0806+2006.}
    \label{fig:sed2}
\end{figure*}

%

%
\end{appendix}
\end{document}